\documentclass[journal]{IEEEtran}
\usepackage[utf8]{inputenc} 
\usepackage[T1]{fontenc}
\usepackage{amsmath,amsfonts}
\usepackage{mleftright}
\usepackage{algpseudocode}
\usepackage{algorithm}
\mleftright
\usepackage{booktabs} 
\usepackage{array}
\usepackage[caption=false,font=normalsize,labelfont=sf,textfont=sf]{subfig}
\usepackage{textcomp}
\usepackage{stfloats}
\usepackage{url}
\usepackage{verbatim}
\usepackage{cite}
\usepackage{etex}
\usepackage{dcolumn}
\usepackage{epsfig}
\usepackage{graphicx}
\usepackage{amssymb}
\usepackage{amsthm}
\usepackage{bm}
\usepackage{enumitem,kantlipsum}
\usepackage{setspace}
\usepackage[pagebackref=false]{hyperref}
\usepackage[svgnames]{xcolor} 
\usepackage{pstricks,pst-node,pst-plot,pstricks-add}
\usepackage{authblk}
\usepackage{algpseudocode}
\usepackage[noabbrev]{cleveref}

\usepackage{tikz}

\usepackage{mathdots}
\usepackage{mathrsfs}
\usepackage{cancel}
\usepackage{color}
\usepackage{siunitx}
\usepackage{array}
\usepackage{multirow}
\usepackage{gensymb}
\usepackage{tabularx}
\usepackage{booktabs}
\usepackage{xspace}

\usepackage[referable]{threeparttablex}

\input{Definition.def}
\graphicspath{{figs/}}

\newcommand{\tabref}[1]{Table~\ref{#1}}
\tikzset{every picture/.style={line width=0.75pt}}
\def\BibTeX{{\rm B\kern-.05em{\sc i\kern-.025em b}\kern-.08em
		T\kern-.1667em\lower.7ex\hbox{E}\kern-.125emX}}
\newcommand{\google}{\texttt{SecAgg}\xspace}
\makeatletter
\newcommand{\algmargin}{\the\ALG@thistlm}
\makeatother
\algnewcommand{\parState}[1]{\State%
    \parbox[t]{\dimexpr\linewidth-\algmargin}{\strut #1\strut}}

\begin{document}
\title{\texttt{ByzSecAgg}: A Byzantine-Resistant Secure Aggregation Scheme for Federated Learning Based on Coded Computing and Vector Commitment}



\author{Tayyebeh Jahani-Nezhad,
        Mohammad~Ali~Maddah-Ali,~\IEEEmembership{Fellow,~IEEE,}
        and~Giuseppe~Caire,~\IEEEmembership{Fellow,~IEEE}
        
\thanks{T.~Jahani-Nezhad and G.~Caire are with the Department of Electrical Engineering and Computer Science, Technische Universität Berlin, 10587 Berlin, Germany (e-mail: {t.jahani.nezhad, caire}@tu-berlin.de).}
\thanks{M.~A.~Maddah-Ali is with 
the Department of Electrical and Computer Engineering, University of Minnesota Twin Cities, MN 55455 USA (e-mail: maddah@umn.edu)}
}

	\maketitle



 \begin{abstract} 
In this paper, we propose \texttt{ByzSecAgg}, an efficient secure aggregation scheme for federated learning that is resistant to Byzantine attacks and  privacy leakages. 
Processing individual updates to manage adversarial behavior, while preserving the privacy of the data against colluding nodes, requires some sort of secure secret sharing. However, the communication load for secret sharing of long vectors of updates can be very high. In federated settings, where users are often edge devices with potential bandwidth constraints, excessive communication overhead is undesirable. \texttt{ByzSecAgg} solves this problem by partitioning local updates into smaller sub-vectors and sharing them using ramp secret sharing. However, this sharing method does not admit bilinear computations, such as pairwise distances calculations, which are needed for distance-based outlier-detection algorithms, and effective methods for mitigating Byzantine attacks. 
To overcome this issue, each user runs another round of ramp sharing, with a different embedding of the data in the sharing polynomial. This technique, motivated by ideas from coded computing, enables secure computation of pairwise distance.  In addition, to maintain the integrity and privacy of the local update, \texttt{ByzSecAgg} also uses a vector commitment method, in which the commitment size \emph{remains constant} (i.e., does not increase with the length of the local update), while simultaneously allowing verification of the secret sharing process. In terms of communication load, \texttt{ByzSecAgg} significantly outperforms the related baseline scheme, known as \texttt{BREA}. 
 \end{abstract}


\section{Introduction}
\IEEEPARstart{F}{ederated learning} (FL) is an emerging distributed learning framework that allows a group of distributed users (e.g., mobile devices) to collaboratively train a global model with their local private data, without sharing the data~\cite{mcmahan2017communication,kairouz2019advances,li2020federated}. 
Specifically, in an FL system with a central server and several users, during each training iteration, the server sends the current state of the global model to the users. Receiving the global model, each user then calculates a local update with its local data, and sends the local update to the server. By aggregating the local updates, the server can update the global model for the next iteration. 
While the local datasets are not directly shared with the server, several studies have shown that a curious server can launch model inversion attacks to reveal information about the training data of the individual users from their local updates~\cite{zhu2020deep,geiping2020inverting}. Therefore, the key challenge to protect users' data privacy is to design \emph{secure aggregation} protocols, which allow the aggregation of the local updates to be computed without revealing each individual data. Moreover, as some users may randomly drop out of the aggregation process (due to low batteries or unstable connections), the server should be able to robustly recover the aggregated local updates of the surviving users in a privacy-preserving manner.

As such motivated, a secure aggregation protocol \google is proposed in~\cite{bonawitz2017practical}. In \google, the local updates are masked before being sent to the server, using private and shared random vectors, such that the shared parts can be canceled out when aggregated. 
One of the major challenges for \google is the communication load, which grows quadratically with the number of users.
There has been a series of works aiming to improve the communication efficiency of \google~(see, e.g.,~\cite{so2021turbo,bell2020secure,choi2020communication,zhao2021information,schlegel2021codedpaddedfl,yang2021lightsecagg,jahani2022swiftagg}). For instance, \texttt{SwiftAgg+}, recently proposed in \cite{jahani2022swiftagg+}, significantly reduces the communication overheads without any compromise on worst-case information-theoretic security, and achieves optimal communication loads within diminishing gaps for secure aggregation.

In~\cite{so2021turbo,bell2020secure,choi2020communication,zhao2021information,schlegel2021codedpaddedfl,yang2021lightsecagg,jahani2022swiftagg,jahani2022swiftagg+}, the \emph{honest-but-curious} model for secure aggregation is considered in which users correctly follow the protocol but some of them may try to gather information about other users' private information, potentially through collusion. However, some studies have also addressed the \emph{malicious} model in which Byzantine adversaries have more capabilities and may run poisoning attacks, i.e., manipulate their inputs to change the outcome of the aggregation~\cite{shen2016auror,bagdasaryan2020backdoor,9945997}. The problem of robustness against Byzantine adversaries in distributed and federated learning is well-researched in two categories: distance-based methods \cite{blanchard2017machine,guerraoui2018hidden,fung2018mitigating,yin2018byzantine} and validation data-based methods \cite{xie2019zeno,xie2020zeno++}. However, those defenses come with some potential privacy issues, as they require the server to process the local model updates. 

As a Byzantine resistant secure aggregation scheme, in \cite{so2020byzantine}, \texttt{BREA}  is proposed  in which
users share their model updates with others using Shamir's secret sharing and then calculate the pairwise distance between the received shares. The server then uses this information to identify and exclude Byzantine users from the aggregation process. \texttt{BREA} suffers from two major shortcomings: (1) to verify secret sharing, in  \texttt{BREA} every single element of data is committed separately, (2) the secret sharing is not designed efficiently. As a result, the loads of commitment and sharing grow linearly with the \emph{aggregated size} of all \emph{local updates}. 
An alternative approach is proposed in \cite{zhang2021safelearning}, which  involves grouping users anonymously and randomly into subgroups with a hierarchical tree structure. The aggregation method similar to \texttt{SecAgg} is performed within each subgroup, and then the process repeats at the next level of subgroups toward the root. The anonymity and randomness of the grouping reduce the probability of successful Byzantine attacks, as attackers do not know which subgroup the compromised device belongs to. However, in that approach, smaller subgroups result in reduced privacy for participants, while larger subgroups make it easier for the Byzantine attacker to hide their malicious model among the honest ones. In \cite{velicheti2021secure}, a similar idea is proposed, which involves repeating the random grouping multiple times to save more updates of benign users through robust aggregation. However, that approach increases the communication loads and also reveals some information about the local updates.
In \cite{rathee2022elsa}, \texttt{ELSA} is proposed, which utilizes a distributed trust setup where two servers in separate trust domains interact and exchange information with one another.
Each iteration begins with two servers selecting a group of users to participate and sharing the current global model with them. These users then use the cryptography module to securely share their local updates with the two servers. The servers then use an interactive protocol to combine these updates and find the aggregation.

 One notable approach that has gained attention is the utilization of concepts from coded computing to enhance the security and efficiency of distributed computing algorithms. Coded computing, originally developed in the context of distributed storage and computation, offers possibilities for optimizing communication and computation tasks in distributed systems \cite{lee2017speeding,yu2017polynomial,yu2020straggler,dutta2019optimal,jahani2021codedsketch,dutta2016short,ferdinand2018hierarchical}. 
Additionally, the concept of secure coded computation has been developed to address the challenge of preserving data privacy in large-scale computations, such as matrix multiplication, ensuring that sensitive information remains protected during computation \cite{chang2018capacity,jia2021capacity,d2020gasp,tang2022adaptive,akbari2021secure}.
 Furthermore, by leveraging ideas from error-correcting codes and network coding, coded computing techniques enable the efficient processing of data and computation while mitigating the impact of Byzantine behaviors in computations \cite{najarkolaei2020coded,yu2019lagrange,soleymani2021list}.

 In this paper, we propose \texttt{ByzSecAgg}, a novel single-server Byzantine-robust secure aggregation scheme in a federated setting. In addition to providing Byzantine-robustness and privacy preservation, \texttt{ByzSecAgg} addresses the issue of high communication load. This is particularly crucial in federated learning settings, where users are edge devices such as mobile phones. It is common for these devices to have bandwidth limitations or be unable to handle high communication loads. \texttt{ByzSecAgg} draws inspiration from the integration of techniques from diverse fields, including coded computing, cryptography and outlier detection algorithms. The proposed scheme is robust against user dropouts, collusion and Byzantine adversarial attacks, and it involves the following steps:
\begin{itemize}[leftmargin=*]
    \item Each user partitions its local update vector into smaller sub-vectors and broadcasts constant-size commitments of them, regardless of the size of the local update. This ensures that the sub-vectors can be proven to be unchanged during the scheme, while still keeping them hidden.
    \item Users securely share these sub-vectors with others using ramp secret sharing, and these shares can be verified using the commitments.
    \item Inspired by coded computing techniques, each user creates another polynomial function to send additional shares of their sub-vectors to other users, allowing for the computation of pairwise distance of shares.
    \item The server then uses these pairwise distances of shares to decode the pairwise distances of the true local updates. Using these distances, the server employs a distance-based outlier detection algorithm to select a subset of users for aggregation.
    \item Finally, the server obtains the aggregation of local updates of the selected users by communicating with the users who locally aggregate the secret shares of the model updates that belong to the selected users.
\end{itemize}
\texttt{ByzSecAgg} ensures the privacy of individual local models by performing computations using secret shares, which prevent users from learning the true values of local updates. Additionally, the server is not able to obtain any information about the local models beyond the aggregation and the pairwise distances which are strictly required by the outlier detection method.
Furthermore, the commitments in \texttt{ByzSecAgg} are binding and computationally hiding.  This means that users cannot obtain any information from the commitments, but they are still able to verify the authenticity of the messages received from others and the validity of the shares. 

\tabref{table} compares the baseline framework \texttt{BREA} in \cite{so2020byzantine} and \texttt{ByzSecAgg} to achieve a certain precision, in terms of communication loads, presented as the number of symbols from the underlying finite field. For a fair comparison, we consider three metrics: the server communication load, the per-user communication load and the commitments size. Server communication indicates the total size of all messages which are sent or received by the server, per-user communication denotes the total size of all messages, that are sent by each user, and the size of the commitments represents the size of commitments made by  users required for message verification.
Compared with the baseline, as shown in \tabref{table}, in the system consisting of $N$ users, \texttt{ByzSecAgg} reduces server and per-user communication loads, as well as significantly decreasing commitment size. The proposed scheme allows for $K$, a design parameter, to be in the range $[1:\frac{N-D+1}{2}-A-T]$, where $D,T$ and $A$ denote the maximum number of dropouts, colluding users, and Byzantine adversaries, respectively, for which the scheme is designed. Based on system parameters $N$ and $L$, which is the size of each local update, the optimal value of $K$ for minimizing communication loads can be chosen.  For instance, for large $L$, in an extreme case, we can choose $K=\mathcal{O}(N)$, which significantly decreases the server and per-user communication loads.
Note that, in the special case when $K=1$, the server communication load in \texttt{ByzSecAgg} is equivalent to that in \texttt{BREA}, and the per-user communication load in \texttt{ByzSecAgg} requires only $N(N-1)$ extra symbols compared to \texttt{BREA} to achieve a higher level of privacy. However, unlike \texttt{BREA}, the commitment size in \texttt{ByzSecAgg} remains constant, regardless of the size of the local updates.

  	\begin{table*}[t]
   \centering
   		\caption{%
		Communication loads of Byzantine-robust secure aggregation frameworks in federated learning.  Here $N$ is the total number of users, $L$ is the size of the local updates, $T$ is the number of colluding users, $D$ is the number of dropouts, $A$ is the number of Byzantine adversaries. In \texttt{ByzSecAgg}, parameter $K\in[1:\frac{N-D+1}{2}-A-T]$ can be chosen based on the network. Note that in this table, we highlighted the parameter $L$ to emphasize the performance of \texttt{ByzSecAgg} in reducing communication load, as $L$ is typically much larger than $N$ in real-world scenarios. 
		}\label{table}
   \resizebox{2\columnwidth}{!}
		 {
			\begin{tabular}{||c |c c c||} 
				\hline
				Approach & Server communication  & Per-user communication  & Commitments size  \\ [0.5ex] 
				\hline\hline
    			\texttt{BREA}~\cite{so2020byzantine} & $(2A+T+1){\textcolor{red}{L}}+(T+A+\frac{1}{2})N(N-1)$ & $N{\textcolor{red}{L}}+\frac{N(N-1)}{2}$ & $TN{\textcolor{red}{L}}$ \\
				\hline
				\texttt{ByzSecAgg}  & $(1+\frac{2A+T}{K}){\textcolor{red}{L}}+(T+A+K-\frac{1}{2}){N(N-1)}$  & $\begin{cases}
				    N{\textcolor{red}{L}}+\frac{3N(N-1)}{2}, & \text{if }K=1\\
                \frac{2N}{K}{\textcolor{red}{L}}+\frac{3N(N-1)}{2}, & \text{if }K>1
				\end{cases}$ & $\begin{cases}
				    N(3T+1),& \text{if }K=1\\
                    N(3K+4T-2),& \text{if }K>1
				\end{cases}$ \\ 
				\hline
			\end{tabular}
		  }
	\end{table*}

\noindent
\textbf{Notation}
For $n\in\mathbb{N}$ the notation $[n]$ represents set $\{1,\dots,n\}$. In addition, for $n_1,n_2\in\mathbb{Z}$ the notation $[n_1:n_2]$ denotes the set $\{n_1,\dots,n_2\}$. 
	Furthermore, the cardinality of set $\mathcal{S}$ is denoted by $|\mathcal{S}|$. In addition, we denote the difference between two sets $\mathcal{A}$ and $\mathcal{B}$ as $\mathcal{A}\backslash\mathcal{B}$, which represents the set of elements belonging to $\mathcal{A}$ but not to $\mathcal{B}$.. In addition, $\mathbb{E}[X]$ and  $H(X)$  refer to the expected value and the entropy of random value $X$ respectively. $\text{Pr}(A)$ is the probability of event $A$.

\section{Problem formulation}
\label{sec:prob}
 		We consider the Byzantine-robust secure aggregation problem, for a federated learning system, consisting of a server and $N$ users. The objective of the server is to train a global model  $\mathbf{w}_g\in\mathbb{R}^{L}$, with dimension $L\in\mathbb{N}$, using the data held at users, by minimizing a global cost function $\mathfrak{L}(\mathbf{w}_g)$.
 	In round $t$ of the training phase, the server broadcasts the global model $\mathbf{w}_g^{(t)}\in\mathbb{R}^{L}$ to all users.
Then each user $n$, $n \in [N]$ computes a private local update $\mathbf{w}_n\in\mathbb{R}^{L}$ based on its private local dataset.
In this paper, we focus on perfect secure aggregation schemes, which rely on operations in a finite field to protect the privacy of the local updates \cite{yang2021lightsecagg,jahani2022swiftagg+,bonawitz2017practical, so2021turbo,bell2020secure,choi2020communication,zhao2021information,schlegel2021codedpaddedfl,jahani2022swiftagg}. Consider that
each user employs an element-wise stochastic quantization method that involves a rounding function $Q:\mathbb{R}\to\mathbb{R}$ 
and a mapping function $\Lambda:\mathbb{R}\to\mathbb{F}$,  which maps the integer numbers to 
elements of a finite field $\mathbb{F}$.
The finite field is selected to be sufficiently large so that during the process of aggregation, there is no risk of encountering the boundary, thereby preventing potential issues.

User $n$ also has a collection of local random variables $\mathcal{Z}_n$, whose elements are selected uniformly at random from $\mathbb{F}^{L}$, and independently of each other and of the local updates.
 It is assumed that each user can directly communicate with the server.
Let $\mathbf{X}^{(L)}_n \in \mathbb{F}^* \cup \{ \perp \}$ denote the message sent by user $n$ to the server.
In addition,  let $\mathbf{M}^{(L)}_{n\to n'} \in \mathbb{F}^* \cup \{ \perp \}$ denote the message that user $n$ sends to user $n'$ in the finite field $\mathbb{F}$ in the algorithm.  The null symbol $\perp$ represents the case where no message is sent. Here $\mathbb{F}^* = \cup_{\ell \in \mathbb{N}} \mathbb{F}^{\ell}$. 
The message $\mathbf{M}^{(L)}_{n\to n'}$ is a function 
	of $\mathbf{w}_n$, $\mathcal{Z}_n$. We denote the corresponding encoding function by $\phi^{(L)}_{n\to n'}$. 
	Similarly, $\mathbf{X}^{(L)}_n$ is a function of $\mathbf{w}_n$, $\mathcal{Z}_n$,  and the messages that user $n$ has received from other users. We denote the corresponding encoding function by $\varphi^{(L)}_{n}$.  Let $\mathcal{U}_{\text{s}}$ refer to a subset of users selected by the server and their local updates are used for aggregation.  Let
 $\mathcal{X}_{\mathcal{S}}=\{\mathbf{X}_n^{(L)}\}_{n\in\mathcal{S}}$ represent the set of messages the server receives from a subset of users $\mathcal{S}$, where $|\mathcal{S}|\le N$. The received messages from the users are decoded by the server using the decoding function $\psi^{(L)}$ in order to use them for the aggregation process.
 In this setting, we assume that a subset of users $\mathcal{D}\subset[N]$ drops out, i.e.,
	stay silent (or send $\perp$ to other users and the server) during the protocol execution. We denote the maximum number of dropped-out users as $D\in\mathbb{N}$.
	We also assume that some of the users are curious and might collude to gain information about the local updates of the other users. Assume that the maximum number of colluding users is denoted by $T\in \mathbb{N}$. Note that the identities of dropouts and colluding users are not known beforehand.
 \subsection{Threat Model} We assume untargeted poisoning attacks, particularly model poisoning attacks which aim to reduce the effectiveness of the global model or prevent its convergence by directly modifying the local updates and selecting malicious parameters before sending them to other nodes (Type-1)~\cite{shen2016auror,bagdasaryan2020backdoor,9945997}. We assume that the attacker can compromise at most $A\in \mathbb{N}$ benign users, and arbitrarily manipulate the local updates sent from these users. We also consider that the attack is a probabilistic polynomial time (PPT) algorithm with respect to a security parameter $\kappa \in \mathbb{N}$~\cite{katz2008introduction}. It means that the attacker can run an algorithm within polynomial time and uses probabilistic methods to try to break the security of the system with a given security parameter $\kappa$. In the following, we refer to these compromised users as Byzantine adversaries. 
 Another adversarial behavior of the Byzantine users is that they might send messages to other nodes, inconsistent with the protocol, requiring verification methods to address this issue (Type-2).
Note that the sets of colluding and adversarial users are not necessarily disjoint, but the problem formulation is stated in the general form.
\subsection{Security Model} 
To mitigate the threat posed by model poisoning attacks, a robust aggregation rules $\Omega:(\mathbb{F}^{*})^{|{\mathcal{S}}|}\to \mathbb{F}^L$ as a defense strategy is employed to address Type-1 adversarial behavior, where $\mathcal{S}$ represents the subset of users that send their messages to the server.  This function represents both the aggregation rule and the user selection method utilized by the server to identify outliers based on the received messages and then eliminate them from the aggregation process.
   The server updates the global model by applying function $\Omega$ to the decoded received messages, as 
 \begin{align*}
     \mathbf{w}_g^{(t+1)} =  \mathbf{w}_g^{(t)}-\delta_t\Lambda^{-1}\bigg(\Omega\big(\psi^{(L)}(\mathbf{X}_n^{(L)},n\in\mathcal{S})\big)\bigg), 
 \end{align*}
 where $\delta_t$ is the learning rate at round $t$ and $\Lambda^{-1}:\mathbb{F}\to\mathbb{R}$ is a demapping function.

In addition, to enable verification of the messages in this adversarial system to protect against Type-2 adversarial behavior, a vector commitment scheme is used. A vector commitment scheme is a cryptographic primitive that enables a user to commit to a vector with the following desirable features. It is computationally infeasible to determine the committed vector from the commitment value. It is also computationally infeasible to find a different vector that maps to the same commitment value. More precisely, a vector commitment scheme \textsf{VC=(Setup,Commit,Witness,Verify)} includes the following components:
 \begin{itemize}[leftmargin=*]
     \item \textsf{Setup}: $1^\kappa \to (\textsf{pp},\textsf{sp})$: this protocol is run by a trusted or distributed authority at the beginning to take security parameter $\kappa$ and generates some public parameters (\textsf{pp}) and some local secret parameters (\textsf{sp}).
     \item \textsf{Commit}:$(\textsf{pp},\bm{\nu})\to C$: this algorithm takes vector $\bm{\nu}$ as input and outputs a commitment $C$. 
     \item \textsf{Witness}: $(\textsf{pp},\bm{\nu})\to {\bm{\omega}}$: this algorithm takes vector $\bm{\nu}$ as input and computes a witness $\bm{\omega}$.
     \item \textsf{Verify}:$(\textsf{pp},C,\bm{\omega})\to b\in \{\texttt{True,False}\} $: this algorithm takes  a witness $\bm{\omega}$ and a commitment $C$ as input and returns either \texttt{True} or \texttt{False} based on the validity of the witness.
 \end{itemize}
 
 \subsection{The Goals}
	 A \emph{Byzantine-robust secure aggregation scheme} consists of the encoding functions  $\phi^{(L)}_{n\to n'}$, $\varphi^{(L)}_{n}$, $n,n' \in [N]$, the decoding function $\psi^{(L)}$, the aggregation rule $\Omega$, and the vector commitment scheme \textsf{VC} such that the following conditions are satisfied:

\noindent	\textbf{1. Correctness:} 
  \begin{itemize}[leftmargin=*]
\item Correctness of the Commitments: For an honest user $n$ with quantized local update $\bar{\mathbf{w}}_n$ of data and a commitment $C_n~=~\textsf{Commit}(\textsf{pp},\bar{\mathbf{w}}_n)$, the created witness $\bm{\omega}_n~=~\textsf{Witness}(\textsf{pp},\bar{\mathbf{w}})$ successfully satisfies $\textsf{Verify}(\textsf{pp},C_n,\bm{\omega}_n)~=~\texttt{True}$.  
     \item Correctness of the Final Result: Despite the presence of at most $D$ dropouts, $A$ Byzantine adversaries, and $T$ colluding users, the server maintains the capability to recover
 $\bar{\mathbf{w}}\triangleq~\frac{1}{|\mathcal{U}_{\text{s}}|}\sum_{n\in\mathcal{U}_{\text{s}}}{\bar{\mathbf{w}}}_{n}$ which must be the outcome of $\Omega\big(\psi^{(L)}(\mathbf{X}_n^{(L)},n\in\mathcal{S})\big)$. 
 \end{itemize}

 \noindent\textbf{2. Robustness against Byzantine Adversaries:}
 \begin{itemize}[leftmargin=*]
   \item Commitment Binding: The algorithm \textsf{Commit} should create a binding and deterministic commitment to the data. Formally, for attacker $\mathcal{A}$ who can simulate any user, it must hold  
     \begin{align*}
         \text{Pr}\bigg( &(\textsf{pp},\textsf{sp})\gets \textsf{Setup}(1^\kappa),
         (\bm{\nu},\bm{\nu}')\gets \mathcal{A}(\textsf{pp},\textsf{sp}):\\
         &\bm{\nu}\ne\bm{\nu}' \wedge \textsf{Commit}(\bm{\nu})=\textsf{Commit}(\bm{\nu}')\bigg)\le \epsilon(\kappa),
     \end{align*}
     where function $\epsilon(.)$ is a \emph{negligible} function, which means for all $c>0$ there exists a $k_c$ such that for all $k>k_c$ we have $\epsilon(k)<\frac{1}{k^c}$. 
     In short, this means that a Byzantine adversary committer can present two distinct values of $\nu$ with the same commitment $C$ with vanishing probability. 
     \item Global Model Resiliency: The aggregation rule $\Omega$ should prioritize global model resiliency, even in the presence of up to $A$ Byzantine adversaries (Type-1), by ensuring that at the end of each iteration, the output of the aggregation rule remains close to the true gradient. Formally, at each iteration of the training process, for the true gradient vector $\mathbf{g}\triangleq\nabla \mathfrak{L}(\mathbf{w}_g^{(t)})$, the output of the aggregation rule, i.e., $\mathbf{g}_{\text{GAR}}\triangleq\Omega\big(\psi^{(L)}(\mathbf{X}_n^{(L)},n\in\mathcal{S})\big)$ must satisfy a well-defined closeness criterion $\xi(\mathbf{g}_{\text{GAR}},\mathbf{g},N,A)$.
     \item Malicious Computation Results: The general scheme should maintain robustness against any malicious computation results produced by Byzantine adversaries (Type-2) that may be sent to the server and other users during each step of the scheme.
 \end{itemize}

\noindent	\textbf{3. Privacy Constraint}: 
 \begin{itemize}[leftmargin=*]
     \item Hiding Property: If a proof $(\textsf{pp},C_n)$ and a witness $\bm{\omega}_n$ for quantized local update $\bar{\mathbf{w}}_n$, $n\in[N]$, are given and the verification \textsf{Verify}(\textsf{pp},$C_n$,$\bm{\omega}_n$) returns \texttt{True},  no user can figure out the value of $\bar{\mathbf{w}}_n$ with non-negligible probability.
     \item Privacy of Individual Local Updates: The privacy constraint ensures that no group of up to $T$ colluding users can extract any information about the local models of other honest users. In addition, after receiving $\mathcal{X}_{\mathcal{S}}$,
     the server should not gain any information about local updates of the honest users, beyond the aggregation of them, and beyond what is strictly required by the user selection (outlier detection) method— specifically, the pairwise distances of local updates used in this work.
 \end{itemize}

For a Byzantine-robust secure aggregation scheme satisfying the above conditions, we define the average per-user communication load and the server communication load as follows: 
	
\begin{definition}[Average per-user communication load] 
The average per-user communication load, denoted by $R_{\text{user}}$,   is defined as the aggregated size of all messages sent by users, i.e.,
		 \begin{align*}
	 	R_{\text{user}}~=~\frac{1}{N} \sum\limits_{\substack{n,n'\in [N],\\n’ \neq n}}\big( H(\mathbf{M}^{(L)}_{n\to n'})+H(\mathbf{X}^{(L)}_{n})\big),
	 	 \end{align*}
     where the base of the $\log(.)$ function in the definition of the entropy function $H(.)$ is the size of the finite field.

\end{definition}
\begin{definition}[Server communication load] 
The server communication load, denoted by $R_{\text{server}}$,  is defined as the aggregated size of all messages received by the server, i.e.,
	 \begin{align*}
	  R_{\text{server}}= \sum_{n\in [N]} H(\mathbf{X}^{(L)}_{n}).
	 \end{align*}
 \end{definition}

In this paper, we propose \texttt{ByzSecAgg}, a scheme for Byzantine-robust secure aggregation in a single server federated setting to meet the aforementioned conditions. 
 In \texttt{ByzSecAgg},  each user partitions its local update vector of length $L$ into $K$ smaller sub-vectors of length $\frac{L}{K}$ and broadcasts constant-size commitments of them. Users then use ramp sharing \cite{blakley1984security} to share the sub-vectors with others, which can be verified using the commitments.  This method allows for a major reduction in communication loads, however, it makes computing the pairwise distances between the local updates and removing the outliers challenging. To address this issue, we use techniques inspired by coded computing, where each user runs another ramp sharing, where data vectors are embedded in the coefficients differently. These two sets of shares admit the computation of the pairwise distances in a very efficient way. The server then receives the pairwise distances of the shares, which are used to recover the pairwise distances of the local updates. These distances are employed in the multi-Krum algorithm \cite{blanchard2017machine} as the outlier detection method, to select $m$ users for aggregation. Finally, communication with the server is required so that the server can obtain the aggregation of local updates of the selected users. To be able to verify that users follow the sharing protocol correctly and also not change their data in the second round of sharing, we need to be able to verify it against some commitment. 
We suggest using some linear commitment scheme \cite{nazirkhanova2021information} that, unlike many existing solutions, the size of the commitment remains constant and does not grow with the size of data.

\section{Preliminaries}
In order to develop a secure aggregation scheme that is resilient to Byzantine adversarial behavior, particularly of Type-1, it is essential to employ a robust aggregation method. Specifically, in this paper, we leverage the multi-Krum algorithm proposed in \cite{blanchard2017machine}. 
Note that any alternative aggregation methods that are robust against Type-1 adversarial attacks and are based on Euclidean pairwise distances of the local updates can also be employed in \texttt{ByzSecAgg}. However, depending on the structure of each method, the proposed scheme might require some modifications.

\subsection{Multi-Krum Algorithm}\label{krum-def}
In the distributed stochastic gradient descent (SGD) problem, and in the presence of Byzantine adversaries (Type-1), using the average of all users’ gradients to update the model parameters is not robust since a single Byzantine user can cause an arbitrarily large error in the update. To handle this, the server must use a gradient aggregation rule (GAR) $\Omega$ that is resistant to malicious gradients that may be produced by up to $A$ Byzantine adversaries. 
At iteration, $t$, assume that each honest user $n$ calculates an estimate $\mathbf{w}_n^{(t)} = \mathbf{G}(\mathbf{w}_g^{(t)}, \zeta_n^{(t)})$ of the gradient of the cost function $\mathfrak{L}$, where $\zeta_n^{(t)}$ is a random variable representing the sample or mini-batch of samples drawn from user $n$'s dataset, and $\mathbf{w}_g^{(t)}$ is the global model parameter received from the server. On the other hand, Byzantine adversarial users may send some arbitrary vector to the server (Type-1 adversarial attack). 
One method to measure resilience against such Byzantine users is through the concept of \emph{$(\gamma,A)$-Byzantine resilience}, as introduced and supported by convergence theories in \cite{blanchard2017machine}. 
\begin{definition}[$(\gamma,A)$-Byzantine Resilience~\cite{blanchard2017machine}]
    Consider $N$ vectors $\mathbf{w}_1,\dots, \mathbf{w}_N\in\mathbb{R}^L$ which are gradient vectors received by the server from $N$ users. If user $i$ is non-Byzantine, then  $\mathbf{w}_i$ is independent identically distributed random vector, $\mathbf{w}_i\sim  \mathbf{G}(\mathbf{w}_g^{(t)}, \zeta_i^{(t)})$, with $\mathbb{E}_{\zeta_i^{(t)}}{\mathbf{G}}= \nabla\mathfrak{L}(\mathbf{w}_g^{(t)})$. Vector $\mathbf{w}_i$ for a Byzantine user can be any arbitrary vector.
     For any angular value $0\le \gamma<\pi/2$ and any integer $0\le A\le N$ denoting the maximum number of Byzantine adversaries, a gradient aggregation rule (GAR) $\Omega$ is called $(\gamma,A)$-Byzantine resilient if vector $\mathbf{w}_{\text{GAR}}\triangleq \Omega(\mathbf{w}_1,\dots,\mathbf{w}_N)$  satisfies the following two conditions:
\begin{enumerate}
    \item $\langle \mathbb{E}[{\mathbf{w}_{\text{GAR}}}],\mathbf{g}\rangle \ge (1-\sin\gamma)\norm{\mathbf{g}}^2>0$, where $\mathbf{g}$ is the true gradient $\mathbf{g}\triangleq\nabla \mathfrak{L}(\mathbf{w}_g^{(t)})$.
    \item For $r\in\{2,3,4\}$, $\mathbb{E}[\norm{\mathbf{w}_{\text{GAR}}}^r]$ is upper bounded by 
    \begin{align*}
        \mathbb{E}[\norm{\mathbf{w}_{\text{GAR}}}^r]\le c\hspace{-2mm}\sum\limits_{r_1+\dots+r_{N-A}=r}\mathbb{E}[\norm{\mathbf{G}}^{r_1}] \dots\mathbb{E}[\norm{\mathbf{G}}^{r_{N-A}}],
    \end{align*}
     where $c$ is a generic constant and $r_1,\dots,r_{N-A}$ are non-negative integers.
\end{enumerate}
\end{definition}
In this definition, Condition 1 ensures that the angle between the true gradient $\mathbf{g}$ and the output $\mathbf{w}_{\text{GAR}}$ is small enough and on average the output is in the same direction as the true gradient. Condition 2 ensures that the second, third, and fourth-order moments of the output of the aggregation rule are bounded by a linear combination of terms $\mathbb{E}[\norm{\mathbf{G}}^{r_1}] \dots\mathbb{E}[\norm{\mathbf{G}}^{r_{N-A}}]$. This is generally necessary to ensure the convergence of the SGD algorithm.

Krum algorithm \cite{blanchard2017machine} is a gradient aggregation rule that is resilient to the presence of $A$ Byzantine adversaries in a distributed system consisting of $N$ users as long as $N>2A+2$. The Krum aggregation algorithm returns the gradient computed by the user with the lowest score, which is determined by considering the $N-A-2$ closest gradients to that user's gradient. More precisely, let $\mathbf{w}_1, \dots , \mathbf{w}_N$ be the gradients received by the server. Krum algorithm assigns to each vector $\mathbf{w}_i$  score
    $s(i)\triangleq \sum\limits_{j: i\to j} \norm {\mathbf{w}_i-\mathbf{w}_j}^2$,
where $i\to j$ denotes that $\mathbf{w}_j$ belongs to the $N-A-2$ closest vector (in terms of squared distance) to $\mathbf{w}_i$ for any $i\ne j$. The output of Krum algorithm is $\text{KR}(\mathbf{w}_1,\dots , \mathbf{w}_N)=\mathbf{w}_{i^*}$, where $i^*$ is the gradient with the lowest score, i.e., for all $i$ we have $s(i^*)\le s(i)$. 

\begin{lemma}[\cite{blanchard2017machine}]
    Consider i.i.d. random vectors $\mathbf{w}_1,\dots, \mathbf{w}_N\in\mathbb{R}^L$ such that $\mathbf{w}_i\sim \mathbf{G}(\mathbf{w}_g, \zeta_i)$, with $\mathbb{E}{\mathbf{G}(\mathbf{w}_g, \zeta_i)}=\nabla \mathfrak{L}(\mathbf{w}_g)$. Define  $\sigma^2(\mathbf{w}_g)\triangleq\frac{1}{L}\mathbb{E}[\norm{\mathbf{G}(\mathbf{w}_g, \zeta_i)-\nabla \mathfrak{L}(\mathbf{w}_g)}^2]$. Let $\tilde{\mathbf{w}}_1,\dots, \tilde{\mathbf{w}}_A$ be any $A$ random vectors that may depend on the $\mathbf{w}_i$'s. If  $N>2A+2$ and  $\eta(N,A)\sqrt{L}\sigma<\norm{\nabla \mathfrak{L}(\mathbf{w}_g)}$, where $\eta(N,A)$ is defined as 
     \begin{align*}
     \eta(N,A)\hspace{-1mm}\triangleq\hspace{-1mm}\sqrt{2\bigg(N-A+\frac{A(N-A-2)+A^2(N-A-1)}{N-2A-2}\bigg)},
 \end{align*}
 then the Krum algorithm is $(\gamma,A)$-Byzantine resilient, where $0\le \gamma<\frac{\pi}{2}$ is defined by $\sin \gamma=\frac{{\eta(N,A)\sqrt{L}\sigma}}{\norm{\nabla \mathfrak{L}(\mathbf{w}_g)}}$. 
\end{lemma}

 The convergence analysis of the SGD using Krum algorithm is presented in  \cite{blanchard2017machine}.
  The proof relies on certain conditions regarding the learning rates and the gradient estimator.  In addition, in that paper, a stronger variant of Krum algorithm is proposed called \emph{multi-Krum}. In multi-Krum, $m\in [N]$, $m<N-2A-2$, gradient vectors $\mathbf{w}_{i^*_1},\dots ,\mathbf{w}_{i^*_m}$ are selected that have the lowest scores and the output of the algorithm is the average of the selected vectors, i.e., $\frac{1}{m}\sum_{k=1}^m \mathbf{w}_{i^*_k}$.

 Note that although multi-Krum offers theoretical robustness against Type-1 adversarial attacks,  it, like other Byzantine-robust methods, remains vulnerable in certain scenarios. These include sophisticated poisoning strategies proposed in \cite{fang2020local,pmlr-v115-xie20a}, attacks tailored for non-IID data distributions as discussed in \cite{huang2023multi}, and attacks based on techniques such as Projected Gradient Descent (PGD), as highlighted in \cite{wang2020attack}. However, these attacks fall beyond the scope and assumptions of this paper.

\section{The Proposed Scheme}\label{proposed_scheme}
In this section, we explain \texttt{ByzSecAgg} in detail.
Consider a federated learning system with one server and $N$ users. The scheme is designed to handle a maximum number of $T$ colluding users, a maximum number of $D$ dropout users, and a maximum number of $A$  Byzantine users, as defined in Section~\ref{sec:prob}.
The adversaries are assumed to be probabilistic polynomial time (PPT) algorithms with respect to the security parameter $\kappa$. 
The sets of colluding users, dropouts, and adversaries are not known beforehand.

User $n$ has a local update $\mathbf{w}_n\in\mathbb{R}^{L}$. 
 We focus on perfect secure aggregation schemes, which rely on operations in a finite field to protect the privacy of the local updates \cite{yang2021lightsecagg,jahani2022swiftagg+,bonawitz2017practical, so2021turbo,bell2020secure,choi2020communication,zhao2021information,schlegel2021codedpaddedfl,jahani2022swiftagg}.
We choose a finite field, $GF(p)$ denoted by $\mathbb{F}_p$, for some prime number $p$ which is large enough.  User $n$ samples  vectors ${\mathcal{Z}}_n=\{{\mathbf{z}}_{n,j},j\in[T]\}$ and $\tilde{\mathcal{Z}}_n=\{{\bar{\mathbf{z}}}_{n,j}, j\in[T]\}$ uniformly at random  from ${\mathbb{F}}^{\frac{L}{K}}_p$, for some parameter $K\in\mathbb{N}$. In addition, user $n$ samples random scalars ${{\mathcal{R}}}_n=\{{r}^{(j)}_{n,i},i\in[2(K+T)-1]\backslash\{K\},j\in[N]\}$ uniformly at random from ${\mathbb{F}}_p$.  \\
 Each user takes the following steps:
 \begin{enumerate}[wide, labelwidth=!, labelindent=0pt]
 \item {\bf Quantization:} User $n$ converts its local update vector $\mathbf{w}_n\in\mathbb{R}^L$ in real numbers to vector $\bar{\mathbf{w}}_n\in\mathbb{F}_p^L$ in  finite field. This conversion allows for the use of finite field operations, which play a crucial role in protecting the privacy of the local updates.
 To achieve this, each user first applies the stochastic rounding function in \cite{gupta2015deep,so2020byzantine} element-wise as follows
\begin{align}\label{rounding}
    Q_q(x)=\begin{cases}
        \frac{\lfloor qx\rfloor}{q}&\text{with probability } 1-(qx-\lfloor qx \rfloor),\\
         \frac{\lfloor qx\rfloor+1}{q} &\text{with probability } qx-\lfloor qx \rfloor,
    \end{cases}
\end{align}
where integer $q\ge 1$ is the number of quantization levels and $\lfloor x\rfloor$  is the largest integer that is less than or equal to $x$. Note that the rounding function is unbiased, i.e., $\mathbb{E}[Q_q(x)]=x$. 
In addition, to represent negative values in the finite field, a mapping function is needed. The quantized version of the local update of user $n$ denoted by $\bar{\mathbf{w}}_n$ is defined as follows
\begin{align}\label{quantization}
    \bar{\mathbf{w}}_n\triangleq \Lambda{\bigg(q Q_q(\mathbf{w}_n)\bigg)},
\end{align}
where $\Lambda:\mathbb{R}\to\mathbb{F}_p$ is a mapping function that is applied element-wise and defined as
\begin{align}
    \Lambda(x)=\begin{cases}
        x&\text{if } x\ge 0,\\
        x+p&\text{if }x<0.
    \end{cases}
\end{align}
It should be noted that in this step, any rounding function that ensures the convergence of the model can be utilized. There are no restrictions on selecting the rounding function in the proposed method.
     \item {\bf Partitioning the local updates: }
     User $n$ partitions its quantized local update $\bar{\mathbf{w}}_n$ into $K\in\mathbb{Z}$ sub-vectors, i.e.,
		  \begin{align*}
		   \bar{\mathbf{w}}_n=[\bar{\mathbf{w}}_{n,1},\bar{\mathbf{w}}_{n,2},\dots,\bar{\mathbf{w}}_{n,K}]^T,
		  \end{align*}
		where each part $\bar{\mathbf{w}}_{n,k}, k\in[K]$ is a vector of size $\frac{L}{K}$ and $K\in[1:\frac{(N-D+1)}{2}-A-T] $. If the value of $K$ does not divide $L$, we can zero-pad the quantized local models.
  \item {\bf Broadcasting the Commitments:}
    Since all steps are done in the presence of adversarial behavior of Type-2,  some initial hiding commitments are needed. This enables each user to commit to certain values without revealing any information about those values. Using the commitments, each user can verify that the messages being received are effectively the ones for which the commitment was created and ensure that all users follow the protocol honestly.
    Inspired by the commitment scheme in \cite{nazirkhanova2021information}, we consider a scheme where each user $n$ produces the commitments as follows 
    \begin{align}\label{commitments}
   {C}_{i}^{(n)}\hspace{-1mm}=
    \begin{cases}
    \prod\limits_{j=1}^{\frac{L}{K}} (g^{\beta^{j-1}})^{[\bar{\mathbf{w}}_{n,i}]_j}, & \hspace{-2mm}\text{if } i\in[K],\\
    \prod\limits_{j=1}^{\frac{L}{K}} (g^{\beta^{j-1}})^{[\mathbf{z}_{n,i-K}]_j}, & \hspace{-2mm}\text{if } i\in[K+1:K+T], \\
     \prod\limits_{j=1}^{\frac{L}{K}} (g^{\beta^{j-1}})^{[\tilde{\mathbf{z}}_{n,i-K-T}]_j}, & \hspace{-2mm}\text{if } i\in[K+T+1:K+2T],\\
    \prod\limits_{j=1}^{N}(g^{\beta^{j-1}})^{r_{n,i-K-2T}^{(j)}},&\hspace{-2mm}\substack{\text{if } i\in[K+2T+1:3K+4T-1]\\ i\ne 2K+2T},
    \end{cases}
        \end{align}
where ${[\bar{\mathbf{w}}_{n,i}]_j}$ is the $j$-th entry of $\bar{\mathbf{w}}_{n,i}$, and
$g$ is a generator of cyclic group $\mathbb{G}$ with product operation, of prime order $p\ge 2^{2\kappa}$. In addition, $\beta\in\mathbb{F}_p$ is a secret parameter generated by a trusted authority. The trusted authority then generates public parameters $(g^{\beta^0},g^{\beta^1},\dots,g^{\beta^{\max{(\frac{L}{K},N)-1}}})$. Recall that the size of each commitment is equal to a single group element. 
 These commitments are binding, and computationally hiding under the assumption that the discrete logarithm problem is hard in $\mathbb{G}$ (see proofs in Subsection \ref{analysis}). 
  \item {\bf Secret Sharing (First Round):}
		User $n$ forms the following polynomial function.
		\begin{align}\label{Share1}
		\mathbf{F}_{n}(x)=\sum\limits_{k=1}^K\bar{\mathbf{w}}_{n,k}x^{k-1}+\sum\limits_{t=1}^T\mathbf{z}_{n,t}x^{K+t-1},
		\end{align}
		where the coefficient of $K$ first terms are the partitions of  $\bar{\mathbf{w}}_{n,k}$, for $k\in[K]$. Each user $n$ uses its polynomial function $\mathbf{F}_{n}(.)$ to securely share its local update with other users.
  Let $\{\alpha_i \in \mathbb{F}_p : i \in [N]\}$ be a set of $N$ distinct non-zero values in $\mathbb{F}_p$. This set is revealed to all users, such that each user $n$ sends one valuation of its created polynomial function at $\alpha_{\tilde{n}}$ to user $\tilde{n}$, for $\tilde{n}\in[N]$. In particular, user $n$ sends a vector
		$\mathbf{s}_{n,\tilde{n}}\triangleq\mathbf{F}_{n}(\alpha_{\tilde{n}})$ to user ${\tilde{n}}$, and each vector has a size of $\frac{L}{K}$. 
	In this step, if a user $m$ drops out and stays silent, $\mathbf{F}_{m}(.)$ are just presumed to be $ \perp$. Since the proposed secret sharing is based on the ramp sharing scheme, the local update is kept private against $T$ colluding users by adding $T$ independent random vectors in \eqref{Share1} to the message  \cite{blakley1984security,jahani2022swiftagg+}.

\item{\bf Verification (First Round):} 
Having received the shares $\mathbf{s}_{\tilde{n},n}$ from users $\tilde{n}\in[N]$, user $n$ can verify them. Since
 each evaluation of polynomial function $\mathbf{F}_{\tilde{n}}(x)$ is a linear combination (coded version) of its coefficients, due to the linear homomorphism, the verification of $\mathbf{s}_{\tilde{n},n}$ can be done using the encoding of the commitments $C_i^{(\tilde{n})}$ as follows
\begin{align}\label{verification}
\prod\limits_{j=1}^{\frac{L}{K}} (g^{\beta^{j-1}})^{[\mathbf{s}_{\tilde{n},n}]_j} \stackrel{\text{?}}{=} \prod\limits_{j=1}^{K+T} (C_j^{(\tilde{n})})^{\alpha_n^{j-1}}.
\end{align}
Using this verification, user $n$ ensures that it receives a valid evaluation of polynomial $\mathbf{F}_{\tilde{n}}(.)$ in \eqref{Share1}. 

\item {\bf Secret Sharing (Second Round):}
   In this step, if $K\ge 2$, user $n$ forms the polynomial function
  \begin{align}\label{share2}
    \tilde{\mathbf{F}}_{n}(x)=\sum_{k=1}^K{\bar{\mathbf{w}}}_{n,k}x^{K-k}+\sum_{t=1}^T
	\tilde{\mathbf{z}}_{n,t}x^{K+t-1},
  \end{align}
  and sends $\tilde{\mathbf{s}}_{n,\tilde{n}}\triangleq\tilde{\mathbf{F}}_{n}(\alpha_{\tilde{n}})$ to user $\tilde{n}$ for $n,\tilde{n}\in[N]$. In addition, user $n$ creates the scalar polynomial function 
  \begin{align}\label{share2-noise}
    N^{(j)}_n(x)=\sum_{\substack{i=1, i\ne K}}^{2(K+T)-1}{r}^{(j)}_{n,i}x^{i-1}, \hspace{1mm}\text{for }  j\in[N]\backslash\{n\},
  \end{align}
 and sends $N^{(j)}_{n,\tilde{n}}\triangleq N^{(j)}_n(\alpha_{\tilde{n}})$ to user $\tilde{n}$ for $j\in[N]\backslash\{n\}$. The coefficient of $x^{K-1}$ in polynomial $N^{(j)}_n(x)$ is equal to zero.
  
  In this step, each user communicates with other users and sends a vector of size $\frac{L}{K}$ and $N-1$ scalar values. It is worth noting that the structure of the created polynomial function $\tilde{\mathbf{F}}_n(x)$ in this step is different from the polynomial in Step~4.\\
    Note that for $K=1$, in the second round of secret sharing, sending only the shares from the scalar polynomial function in \eqref{share2-noise} is sufficient.

\item {\bf Verification (Second Round):} To ensure that user $n$ receives a valid evaluation of polynomials $\tilde{\mathbf{F}}_{\tilde{n}
}(x)$ and $N_{\tilde{n}}^{(j)}(x)$ from user $\tilde{n}$, it can check
\begin{align}\label{verification2}
\prod\limits_{j=1}^{\frac{L}{K}} (g^{\beta^{j-1}})^{[\tilde{\mathbf{s}}_{\tilde{n},n}]_j} \stackrel{\text{?}}{=} \prod\limits_{j=1}^{K} (C_j^{(\tilde{n})})^{\alpha_n^{K-j}}\prod\limits_{i=K+T+1}^{K+2T} (C_i^{(\tilde{n})})^{\alpha_n^{i-T-1}},\\\nonumber
\prod\limits_{j=1}^{N} (g^{\beta^{j-1}})^{N^{(j)}_{\tilde{n},n}} \stackrel{\text{?}}{=} \prod\limits_{\substack{i=K+2T+1,\\ i\ne 2K+2T}}^{3K+4T-1} (C_i^{(\tilde{n})})^{\alpha_n^{i-K-2T-1}},
\end{align}
using the available commitments. In this verification, user $n$ can not only confirm that it receives a valid evaluation of polynomial ${\tilde{\mathbf{F}}}_{\tilde{n}}(x)$ from user $\Tilde{n}$, but it can also ensure that user $\tilde{n}$ correctly creates polynomial ${\tilde{\mathbf{F}}}_{\tilde{n}}(x)$ without any malicious behavior. This is because the initial commitments made in Step~3 are utilized for the verification. 
 \item {\bf Computing Noisy Inner Products of Shares:}
		In this step,  user $n$ calculates the following inner product and sends the result to the server.
 \begin{align}\label{distances}
  \bar{d}_{i,j}^{(n)}=\langle\mathbf{F}_i(\alpha_n)-\mathbf{F}_j(\alpha_n),\tilde{\mathbf{F}}_i(\alpha_n)-\tilde{\mathbf{F}}_j(\alpha_n)\rangle+ N^{(j)}_{i,n}+N^{(i)}_{j,n},
  \end{align}
 where $i,j\in[N]$ and $i<j$.  
Here, one can see that the inner product of the shares, $\bar{d}_{i,j}^{(n)}$, is an evaluation of a polynomial function
\begin{align}\label{d_ij}
   \bar{d}_{i,j}(x)\triangleq \sum_{\ell=0}^{2(K+T-1)}a_{\ell;i,j}x^{\ell}+N^{(j)}_{i}(x)+N^{(i)}_j(x), 
\end{align}
 at point $\alpha_n$, where  $a_{\ell,i,j}$ is the coefficient of $x^{\ell}$. In this expansion, it can be shown that the coefficient of $x^{K-1}$ is 
 \begin{align*}
     a_{K-1,i,j}=\sum_{k=1}^{K}\norm{\bar{\mathbf{w}}_{i,k}-\bar{\mathbf{w}}_{j,k}}^2.
  \end{align*}
 For $K=1$, user $n$ computes $\bar{d}_{i,j}^{(n)}=\norm{\mathbf{F}_i(\alpha_n)-\mathbf{F}_j(\alpha_n)}^2+N^{(j)}_{i,n}+N^{(i)}_{j,n}$ and sends the result to the server.
\item {\bf Distance Recovery at the Server:} Since $\bar{d}_{i,j}(x)$ is a polynomial function of degree $2(K+T-1)$, the server can use  Reed-Solomon decoding \cite{lin2001error} to  recover all the coefficients of this polynomial using $2(K+T)-1$ evaluations of $\bar{d}_{i,j}(x)$. Since in the setting, there are at most $A$ adversarial users (Type-2),  $\bar{d}_{i,j}(x)$ can only be correctly recovered if the server receives at least $2(K+T+A)-1$ outcomes from the users.
In addition, $\bar{d}_{i,j}^{(n)}$, for  that are sent to the server from the non-adversarial users $n\in[N]\backslash\mathcal{D}$ are indeed equal to $\bar{d}_{i,j}(\alpha_n)$, for $i<j\in[N]$.
Therefore, the server is able to recover $\bar{d}_{i,j}(x)$. The coefficient of $x^{K-1}$ in $\bar{d}_{i,j}(x)$ is equal to $  a_{K-1;i,j}=\norm{\bar{\mathbf{w}}_{i}-\bar{\mathbf{w}}_{j}}^2$ that the server looks for in order to find the outliers. According to \eqref{distances},  the remaining recovered coefficients of $\bar{d}_{i,j}(x)$ are distorted by noise and do not disclose any additional details about the local updates apart from the targeted pairwise distance sought by the server.

Then the server converts the calculated distances from the finite field to the real domain as follows.
\begin{align}\label{recovery}
    d_{i,j}=\frac{\Lambda^{-1}(a_{K-1;i,j})}{q^2},
\end{align}
where $\Lambda^{-1}:\mathbb{F}_p\to \mathbb{R}$ is a demapping function which is applied element-wise and is defined as 
\begin{align}\label{demapping}
    \Lambda^{-1}(\bar{x})=\begin{cases}
        \bar{x} & \text{if } 0\le \bar{x}<\frac{p-1}{2},\\
        \bar{x}-p& \text{if } \frac{p-1}{2}\le \bar{x}<p.
    \end{cases}
\end{align}
Assuming a sufficiently large field size $p$, we can guarantee the accurate recovery of pair-wise distances. 
\item {\bf Outlier Detection at the Server:}  In this step, the multi-Krum algorithm (see \cite{blanchard2017machine} and Subsection \ref{krum-def}), a distance-based outlier detection method (or closeness criterion), is employed to ensure that the local updates selected by the server are consistent with each other. This will help to eliminate any local updates that are significantly different from the others (adversarial behavior of Type-1) and may not be suitable for inclusion in the update. 
Upon completion of the algorithm, the server selects a group of $m$ users,  whose local updates are close to each other and broadcasts a list of the chosen users, denoted by $\mathcal{U}_{\text{s}}\subset [N]$. As stated in \cite{blanchard2017machine}, the multi-Krum algorithm ensures the resiliency of the global model, as defined in Section \ref{sec:prob}, and the convergence of the sequence of gradients (See Subsection \ref{analysis}).
\item{\bf Aggregation of the Shares:}
Each user calculates the aggregation of shares of users belonging to set $\mathcal{U}_{\text{s}}$, i.e., $\mathbf{s}_n~=\sum_{\tilde{n}\in\mathcal{U}_{\text{s}}}\mathbf{F}_{\tilde{n}}(\alpha_n)$, and sends it to the server. In this step, each user sends a vector of size $\frac{L}{K}$ to the server.
\item {\bf Recovering the Aggregation:} 
Let us define
\begin{align}\label{final_poly}
	\mathbf{F}(x)&\triangleq\sum_{n\in\mathcal{U}_{\text{s}}}\mathbf{F}_{n}(x)\\\nonumber
 &=\sum\limits_{k=1}^K x^{k-1}\sum_{n\in\mathcal{U}_{\text{s}}}\bar{\mathbf{w}}_{n,k}
 +\sum\limits_{t=1}^T x^{K+t-1}\sum_{n\in\mathcal{U}_{\text{s}}} \mathbf{z}_{n,t},
		\end{align}
as a polynomial of degree $K+T-1$. Let  $\mathcal{X}_{\mathcal{S}}\triangleq\{\mathbf{s}_n\}_{n\in\mathcal{S}}$ denote the set of messages received by the server, where $\mathcal{S}=[N]\backslash\mathcal{D}$. We note that $\mathbf{s}_n$, for non-Byzantine users, is the evaluation of $\mathbf{F}(x)$ at $\alpha_n$. Thus, if the server receives a minimum of $K+T+2A$ results from the users, it can accurately recover the polynomial function $\mathbf{F}(x)$ using Reed-Solomon decoding. In this decoding process, the first $K$ terms correspond to the partitions of $\bar{\mathbf{w}}=\sum_{n\in\mathcal{U}_{\text{s}}}\bar{\mathbf{w}}_n$. To ensure successful aggregation, the field size $p$ must be sufficiently large to avoid encountering boundary issues during the process.

After recovering the aggregate of the users' local updates, the server updates the global model for the next iteration using the following procedure.
\begin{align}\label{update}
    \mathbf{w}^{(t+1)}_g =\mathbf{w}^{(t)}_g-\frac{\delta_t}{q|\mathcal{U}_s|}\Lambda^{-1}\big(\bar{\mathbf{w}}\big),
\end{align}
where demapping function $\Lambda^{-1}(.)$ is defined in \eqref{demapping}, and $\delta_t$ is the learning rate at round $t$. 

Algorithm \ref{alg} summarizes \texttt{ByzSecAgg} based on the aforementioned steps.
\begin{algorithm}
\caption{The proposed Byzantine-Resistant Secure Aggregation Scheme: \texttt{ByzSecAgg}}\label{alg}
\begin{algorithmic}[1]
\For{ each iteration $t=0, \dots, T$} 
\For{each user $n\in[N]$ in parallel}
\parState {Get the global model $\mathbf{w}_g^{(t)}$ from the server}
\parState {%
Compute the local update $\mathbf{w}_n$ based on the local dataset}
\parState {%
Create the quantized update $\bar{\mathbf{w}}_n$ using \eqref{quantization} }
\parState{Partition the local update into $K$ parts}
\parState{%
  Generate commitments $C^{(n)}_i$, $i\in[3K+4T-1]$ using \eqref{commitments} and broadcast them }
\parState{%
  Compute the first-round secret share $\mathbf{s}_{n,\tilde{n}}$ using \eqref{Share1} and send to user $\tilde{n}\in[N]$}
\parState  {Verify the received shares $\{\mathbf{s}_{\tilde{n},n}\}_{\tilde{n}\in[N]}$ using \eqref{verification}}
\parState{Compute the second-round secret shares $\tilde{\mathbf{s}}_{n,\tilde{n}}$ and noise shares $\{N_{n,\tilde{n}}^{(j)}\}_{j\in[N]}$ using \eqref{share2} and \eqref{share2-noise} respectively and send to user $\tilde{n}\in[N]$}
\parState{Verify the received shares $\tilde{\mathbf{s}}_{\tilde{n},n}$ and $\{N_{\tilde{n},n}^{(j)}\}_{j\in[N]}$ by testing \eqref{verification2} }
\parState{Compute values $\bar{d}^{(n)}_{i,j}$, for $i,j\in[N]$, using \eqref{distances} and send to the server}
\EndFor
\parState{Server recovers pairwise distances $\norm{\bar{\mathbf{w}}_i-\bar{\mathbf{w}}_j}^2$, for $i<j\in[N]$ after receiving the results using Reed-Solomon decoding }
\parState{Server converts the calculated distances from the finite field to the real domain and recovers $\{d_{i,j}\}_{i<j\in[N]}$ using \eqref{recovery}. }
\parState{Server selects set $\mathcal{U}_s$ of users by applying multi-Krum algorithm on $\{d_{i,j}\}_{i<j\in[N]}$  }
\parState{Server broadcast set  $\mathcal{U}_s$ to all users}
\For{user $n=1,\dots, N$}
\parState{Calculate the aggregation of the first-round shares of users belonging to set $\mathcal{U}_s$ and send the result to the server}
\EndFor
\parState{Server recovers $\bar{\mathbf{w}}=\sum_{n\in\mathcal{U}_{\text{s}}}\bar{\mathbf{w}}_n$  after receiving sufficient results from users and employing Reed-Solomon decoding}
\parState{Server updates the global model using \eqref{update}}
\EndFor
\end{algorithmic}
\end{algorithm}
\end{enumerate}
   \begin{remark}
     The proposed scheme uses some methods to mitigate adversarial behaviors of Type-1 and Type-2.  To protect against adversarial behavior of Type-1, where attempts are made to manipulate the global model at the server by altering their local updates, \texttt{ByzSecAgg} employs a distance-based outlier detection mechanism (multi-Krum algorithm). This mechanism ensures the robustness of the global model against such adversarial modifications. Additionally, to protect the privacy of local updates in the outlier detection mechanism, \texttt{ByzSecAgg} proposes a privacy-preserving distance computation method, inspired by ramp secret sharing and coded computing. For protection against adversarial behavior of Type-2, first, a vector commitment scheme is utilized in \texttt{ByzSecAgg} to ensure that users follow the protocol correctly when creating secret shares of their local updates in two different rounds. In addition, the ideas from error-correcting codes are used to ensure the correctness of the calculations of users.
 \end{remark}
\begin{remark}
    To eliminate outliers while minimizing communication requirements, \texttt{ByzSecAgg} uses two different rounds of secret sharing to securely calculate the pairwise distance between the local updates. It is worth noting that the second round of secret sharing is designed to only allow the server to calculate the pairwise distance between the local updates, rather than the pairwise distance between the \emph{partitions} of local updates.
\end{remark}
\section{Analysis of the Proposed Scheme}\label{analysis}

\subsection{{Main results}}
In this subsection, we present the main results, achieved by \texttt{ByzSecAgg} described in Section \ref{proposed_scheme}.
 \begin{theorem}  \label{theorem}
     We assume that (i) the cost function $\mathfrak{L}$ is three times differentiable with continuous derivatives, and is non-negative, i.e., $\mathfrak{L}(x)\ge 0$; (ii) the learning rates satisfy $\sum_t\delta_t=\infty$ and $\sum_t\delta_t^2<\infty$; (iii) the second, the third, and the fourth moments of the quantized gradient estimator satisfy $\mathbb{E}_{Q,\zeta}\norm{Q_q(\mathbf{G}(\mathbf{w}_g,\zeta))}^r\le A_r+B_r\norm{\mathbf{w}_g}^r$ for some constant $A_r$ and $B_r$, where $r\in 2,3,4$; (iv) there exist a constant $0\le\gamma<\pi/2$ such that for all $\mathbf{w}_g\in\mathbb{R}^L$, $\eta(N,A)\sqrt{L\sigma^2(\mathbf{w}_g)+\frac{L}{4q^2}}\le \norm{\nabla\mathfrak{L}(\mathbf{w}_g)}\sin \gamma$; (v) the gradient of the cost function $\mathfrak{L}$ satisfies that for $\norm{\mathbf{w}_g}^2\ge R$, there exists constants $\varepsilon>0$ and $0\le \theta<\pi/2-\gamma$ such that
     \begin{align*}
         \norm{\nabla\mathfrak{L}(\mathbf{w}_g)}\ge \varepsilon>0,\\
         \frac{\mathbf{w}_g^T\mathfrak{L}(\mathbf{w}_g)}{\norm{\mathbf{w}_g}\norm{\nabla\mathfrak{L}(\mathbf{w}_g)}}\ge \cos \theta.
     \end{align*}
Then, \texttt{ByzSecAgg} guarantees, 
\begin{itemize}
    \item  Robustness against $D$ dropouts, and $A$ Byzantine adversarial users (Type-1 and Type-2) such that the trained model is $(\gamma,A)$-Byzantine resilient,
    \item The sequence of the gradients $\nabla\mathfrak{L}(\mathbf{w}_g^{(t)})$ converges almost surely to zero, 
    \item  Any group of up to $T$ colluding users cannot extract any information about the local models of other honest users. The server cannot gain any information about local updates of the honest users, beyond the aggregation of them and pair-wise distances. 
\end{itemize}
In addition, the communication loads in \texttt{ByzSecAgg} are as follows
  \begin{align*}
	 R_{\text{server}}&=  \big(1+\frac{2A+T}{K}\big)L+(T+A+K-\frac{1}{2}){N(N-1)},\\ 
	   R_{\text{user}}&\le \min(\frac{2N}{K},N)L+\frac{3N(N-1)}{2},
	  \end{align*}
 symbols from $\mathbb{F}_p$, 
 for some 
$K\in\mathbb{N}$ in $[1:\frac{N-D+1}{2}-A-T]$ 
 $N\ge~2A+D+\max(2K+2T-1,m+3)$ and $m~<~N-2A-D-2$. Furthermore, the size of the commitment for each user in the achievable scheme remains constant, equal to $(3T+3K-2)+\mathbb{I}(K>1)(T)$, regardless of the size of the individual local updates $L$. Additionally, $\mathbb{I}(.)$ is an indicator function.
 \end{theorem}
\begin{remark}
  As explained in Section \ref{proposed_scheme}, $K$ and $m$ are designed parameters. $K$ is the number of partitions of the local updates. By changing $K$, one can change  $R_{\text{server}}$ and  $R_{\text{user}}$. One option is to choose $K$ such that the aggregation of the per-user communication load and the server communication load is minimized. On the other hand,
 $m$ is the number of users whose local updates are selected by the server for aggregation, as discussed in Subsection \ref{krum-def} and Section \ref{proposed_scheme}. 
 \end{remark}
	
\begin{proof}\label{proof_comm}
In the following, we prove that \texttt{ByzSecAgg} satisfies the following conditions:\\
  \noindent\textbf{1) Correctness:}
     \begin{itemize}[leftmargin=*]
     \item Correctness of Commitments:
     The function $h(x)\triangleq g^x$, in \eqref{commitments}, creates a bijection mapping between the finite field $\mathbb{F}_p$ and cyclic group $\mathbb{G}$, where $g$ is a generator of $\mathbb{G}$. This function also has the linear homomorphic property which states that $\forall a_1,\dots,a_n\in\mathbb{F}_p$ and $\forall x_1,\dots,x_n\in\mathbb{F}_p$ we have
     \begin{align*}
         h\bigg(\sum_{i=1}^n a_ix_i \bigg)=\prod_{i=1}^nh(x_i)^{a_i}.
     \end{align*}
     If the user is honest, it will share the evaluations of the same polynomial $\mathbf{F}_n(x)$ $\big(\Tilde{\mathbf{F}}_n(x)$ and $N_n^{(j)}(x)\big)$   in the first (second) round of secret sharing that it had initially committed to in Step 3. 
     Due to the linearly homomorphic property of the commitments in \eqref{commitments}, the verification step in \eqref{verification} (in \eqref{verification2}) passes for the shares of the honest user.

\item Correctness of the Final Result: The received vectors by the server at the end of Step 12 are different evaluations of polynomial  $\mathbf{F}(x)$ of degree $K+T-1$ defined in \eqref{final_poly}. The correctness condition of the final result of \texttt{ByzSecAgg} is satisfied as the server can correctly recover $\sum_{n\in\mathcal{U}_{\text{s}}}\bar{\mathbf{w}}_n$ by utilizing the Reed-Solomon decoding after receiving $K+T+2A$ outcomes from the users with at most $A$  adversarial outcomes. 
     \end{itemize}
    \textbf{2) Robustness against Byzantine Adversaries:}
     \begin{itemize}[leftmargin=*]
         \item  Commitment Binding:\
        The Discrete Logarithm (DL) Assumption \cite{kate2010constant} states that given a prime $p$, a generator $g$ of $\mathbb{G}$, and an element $a\in\mathbb{F}_p$, no adversary with a probabilistic polynomial-time algorithm can compute $a$ given $g$ and $g^a$, i.e., $\text{Pr}[\mathcal{A}\big(g,g^a\big)=a]=\epsilon(\kappa)$ (negligible) for any such adversarial attack algorithm $\mathcal{A}$.

In addition, consider the polynomial function $q_{\mathbf{w}_{n,i}}(x)$ defined as 
\begin{align*}
                 {q}_{\mathbf{w}_{n,i}}(x)\triangleq [\bar{\mathbf{w}}_{n,i}]_1+x [\bar{\mathbf{w}}_{n,i}]_2+\dots+x^{\frac{L}{K}-1}[\bar{\mathbf{w}}_{n,i}]_{\frac{L}{K}},
             \end{align*}
where $[\bar{\mathbf{w}}_{n,i}]_j$ is the $j$-th entry of the $i$-th partition of the quantized local update $\bar{\mathbf{w}}_{n}$. It can be verified that commitments $C_i^{(n)}$ in \eqref{commitments} for $i\in[K]$ are equal to $g^{q_{\mathbf{w}_{n,i}}(\beta)}$.

 By contradiction, assume that there exists an adversary $\mathcal{A}$ that breaks the binding property of commitments. This means that the adversary creates two vectors $\bar{\mathbf{w}}_{n,i}$ and $\bar{\bm{\nu}}_{n,i}$, where $\bar{\mathbf{w}}_{n,i}\ne \bar{\bm{\nu}}_{n,i}$ but $C_i^{(n)}\triangleq g^{q_{\mathbf{w}_{n,i}}(\beta)}=g^{q_{\bm{\nu}_{n,i}}(\beta)}$.
 
 In this case, it can be shown that the adversary can break the DL assumption, which means it is able to recover secret $\beta$ by having $g^{\beta}$. Define the polynomial function $\tilde{q}(x)$ as $q_{\mathbf{w}_{n,i}}(x)-q_{\bm{\nu}_{n,i}}(x)$. The corresponding commitment is equal to $\tilde{C}_i^{(n)}\triangleq g^{\tilde{q}(\beta)}=g^{q_{\mathbf{w}_{n,i}}(\beta)}/g^{q_{\bm{\nu}_{n,i}}(\beta)}=1$ because the commitment scheme is homomorphic. Therefore, $\tilde{q}(\beta)=0$, which means that $\beta$ is a root of the polynomial $\tilde{q}(x)$. 
The adversary can easily solve the instance of the discrete logarithm problem and find the secret parameter $\beta$ by factoring $\tilde{q}(x)$ \cite{shoup2009computational}. This implies that breaking the binding property would enable solving a problem that is assumed to be computationally difficult, reinforcing the notion that the commitment scheme must possess the binding property.

For the evaluation binding, we can present a similar argument.  Let's assume there exists an adversarial algorithm $\mathcal{A}$ that breaks the evaluation binding property of commitment $C$ and computes two different evaluations for $\mathbf{F}_n(x)$ in \eqref{Share1} at point $\alpha$ that satisfy \textsf{Verify} in \eqref{verification}. We can construct an algorithm $\mathcal{B}$ that uses $\mathcal{A}$ to break the DL assumption. $\mathcal{B}$ presents a DL instance $(g,g^{\beta},\dots,g^{\beta^{\frac{L}{K}}})$. Algorithm $\mathcal{A}$ outputs commitments $C$ and two distinct values for share, i.e., $\mathbf{F}_n(\alpha)$ and $\mathbf{F}'_n(\alpha)$, that satisfy \eqref{verification}. Specifically, we have 
\begin{align*}
  \prod\limits_{j=1}^{\frac{L}{K}} (g^{\beta^{j-1}})^{[\mathbf{F}_n(\alpha)]_j} = \prod\limits_{j=1}^{\frac{L}{K}} (g^{\beta^{j-1}})^{[\mathbf{F}'_n(\alpha)]_j}=\prod\limits_{j=1}^{K+T} (C_j^{({n})})^{\alpha^{j-1}}.  
\end{align*}
 Consequently, we observe that 
 \begin{align*}
    \prod\limits_{j=1}^{\frac{L}{K}} (g^{\beta^{j-1}})^{[\mathbf{F}_n(\alpha)]_j-[\mathbf{F}'_n(\alpha)]_j}=1.  
 \end{align*}
 This implies that $\beta$ is a root of polynomial ${F}''(x)=\sum_{j=1}^{\frac{L}{K}}x^{j-1}([\mathbf{F}_n(\alpha)]_j-[\mathbf{F}'_n(\alpha)]_j)$. Therefore, algorithm $\mathcal{B}$ can compute $\beta$ by factoring ${F}''(x)$ once it receives the results from $\mathcal{A}$. Thus, the success probability of solving the DL instance is the same as the success probability of $\mathcal{A}$.
        
         \item Malicious Computation Result (Type-2 Adversarial Attack): 
         Here we discuss the adversarial attack of Type-2 and how \texttt{ByzSecAgg} resolves it:
  \begin{itemize}[leftmargin=*]
      \item 
      The transmission of invalid secret shares in Step 4 and Step 6: this kind of attack is thwarted through the utilization of the commitments, in which the validity of secret shares at each round can be confirmed by verifying \eqref{verification} and \eqref{verification2}, provided that $N-D\ge2A+1$.
      \item
      The transmission of incorrect pairwise distance of shares to the server in Step 8 or incorrect aggregation of the shares of selected users in Step 11: Since the values sent at each of these two steps must represent different evaluation points of a certain polynomial, this kind of attack will be detected and corrected based on Reed-Solomon decoding algorithm with at most $D$ erasures and at most $A$ errors, provided that $N-D\ge 2(K+T+A)-1$ in Step 8 and $N-D\ge K+T+2A$ in Step 11.
      \end{itemize}
         \item Global Model Resiliency (Type-1 Adversarial Attack):   Byzantine users have the ability to modify their local updates to manipulate the global model and can send arbitrary vectors as their local updates to the server. Additionally, in line with classical assumptions in machine learning literature, each data sample used for computing the local update is uniformly and independently drawn by honest users. To detect outliers, \texttt{ByzSecAgg} uses the multi-Krum algorithm introduced in Subsection \ref{krum-def}. Instead of the estimator $\mathbf{G}(\mathbf{w}_g^{(t)},\zeta_n^{(t)})$, in the proposed scheme, the local update of the honest user is created using the quantized estimator $Q_q(\mathbf{G}(\mathbf{w}_g^{(t)},\zeta_n^{(t)}))$. In \cite{so2020byzantine}, it is demonstrated that even when the multi-krum algorithm is applied to the quantized vector $Q_q(\mathbf{w}_n)$ (as used in Step 1 of the proposed scheme), it remains $(\alpha,A)$-Byzantine resilient and the sequence of the gradients converges almost surely to zero. This is due to the fact that the quantization is unbiased and has a bounded variance.  According to Subsection \ref{krum-def} if $m<N-2A-D-2$, the multi-Krum algorithm can prevent this type of attack and ensure that the scheme is $(\gamma,A)$-Byzantine resilient and guarantees the convergence. Note that the first five conditions in Theorem~\ref{theorem}, i.e., (i)-(v),  are essential in the convergence proof of multi-Krum algorithm (see \cite{blanchard2017machine,so2020byzantine}).
     \end{itemize}
    \textbf{3) Privacy Constraint:}
     \begin{itemize}
         \item Hiding Property: For the sake of contradiction, suppose there exists an adversarial attack algorithm $\mathcal{A}$ that can break the hiding property of commitment $C$ and correctly compute vector $\bm{\nu}\in\mathbb{F}_p$. We will demonstrate how to utilize $\mathcal{A}$ to construct another algorithm $\mathcal{B}$ that can break the Discrete Logarithm (DL) assumption.\\
Let $(g,g^a)$ be a DL instance that $\mathcal{B}$ aims to solve. $\mathcal{B}$ selects a random $\tilde{\beta}\in\mathbb{F}^L_p$ and computes $(g,g^{\tilde{\beta}},\dots,g^{\tilde{\beta}^{L-1}})$. It then considers a vector $\bm{\nu}=[a,\nu_1,\nu_2,\dots,\nu_L]$, where $\nu_1,\nu_2,\dots,\nu_{L-1}\in\mathbb{F}_p$ are chosen arbitrarily, and the first entry of the vector $a$ is the answer for the DL instance. Since $\mathcal{B}$ knows $g^a$, it can compute the vector commitment as $C=g^a.g^{\sum{i=1}^{L-1}\nu_i\tilde{\beta}^i}$ and sends it to $\mathcal{A}$.
Upon receiving the commitment, $\mathcal{A}$ computes and returns vector $\bm{\nu}$. At this point, $\mathcal{B}$ can extract the solution $a$, as it corresponds to the DL instance. Thus, the success probability of solving the DL instance using $\mathcal{B}$ is equivalent to the success probability of $\mathcal{A}$.
The aforementioned construction illustrates that if an adversary $\mathcal{A}$ can break the hiding property of commitment $C$, then we can utilize it to construct another algorithm $\mathcal{B}$ capable of breaking the DL assumption. 
        \item Privacy of Individual Local Updates:
The hiding property of the commitments ensures that neither the server nor any user can compute the local model $\bar{\mathbf{w}}_n$ (as well as noise vectors $\mathbf{z}_n$ and $\tilde{\mathbf{z}}_n$) from the commitments $C_i^{(n)}$ in \eqref{commitments}. Thus, by considering the messages exchanged during the scheme, it is sufficient to establish the privacy of each local update against a group of up to $T$ colluding users, denoted by set $\mathcal{T}$. Denote the colluding users by $\tilde{U}_1, \tilde{U}_2,\dots,\tilde{U}_T$.   Let us define the set of messages received by $\tilde{U}_i$ as $\mathcal{M}_{\tilde{U}_i}$, which includes two types of messages: $\mathbf{s}_{n,i}$, $\tilde{\mathbf{s}}_{n,i}$ and $N_{n,i}^{(j)}$ for $n\in[N]\setminus\{\mathcal{T}\cup\mathcal{D}\}$, $j\in [N]\setminus\{n\}$. These messages correspond to the shares from the first and second rounds of secret sharing in Step~4 and Step~6, respectively. According to \eqref{Share1} and \eqref{share2}, the fact that the random vectors are chosen uniformly and independently at random from $\mathbb{F}_p^{\frac{L}{K}}$, and directly from the privacy guarantee in ramp secret sharing \cite{blakley1984security}, for $ n\in[N]\setminus \mathcal{T}$,  we have 
\begin{align*}
    &\text{Pr}\big( \bar{\mathbf{w}}_n=\bm{\nu}\big| \mathcal{M}_{\tilde{U}_i}, i\in[T]\big)\\
    &=\text{Pr}\big( \bar{\mathbf{w}}_n=\bm{\nu} \big| \mathbf{s}_{n,i},\tilde{\mathbf{s}}_{n,i}, N_{n,i}^{(j)},j\in[N]\setminus{n},i\in[T]\big)\\
    &=\text{Pr}\big(\bar{\mathbf{w}}_n=\bm{\nu}\big),
\end{align*}
which means that the colluding users cannot gain any further information about the local updates of the other users. 

Furthermore, the server is also provided with $\bar{d}_{i,j}^{(n)}$ for $i<j\in[N]$ and $\mathbf{s}_n=\sum_{\tilde{n}\in\mathcal{U}_s}\mathbf{s}_{\tilde{n},n}$ from users $n\in[N]\setminus\mathcal{D}$, from which it can reconstruct $\bar{d}_{i,j}^{(n)}(x)$ as defined in \eqref{d_ij}, as well as the polynomial function $\mathbf{F}(x)$ as defined in \eqref{final_poly}. Therefore, from $\mathbf{F}(x)$ the server can recover the aggregation $\bar{\mathbf{w}}=\sum_{n\in\mathcal{U}_{\text{s}}}\bar{\mathbf{w}}_n$ and from the coefficient of $x^{K-1}$ in $\bar{d}_{i,j}^{(n)}(x)$ it can recover the mutual distance $\norm{\bar{\mathbf{w}}_{i}-\bar{\mathbf{w}}_{j}}^2$. It is important to note that due to the inclusion of random noises during the computation in Step 8 by each user, the server cannot retrieve any other details about the local models. The only information that can be obtained is the mutual distances between the local updates, which play a vital role in outlier detection.
     \end{itemize}

\subsection{Communication Loads in \texttt{ByzSecAgg}}
	According to Step~12, the total number of vectors that need to be received by the server to recover the final aggregation is $T+K+2A$, each of size $\frac{L}{K}$ symbols. Moreover, in order to recover the pairwise distance $d_{i,j}$ between $\bar{\mathbf{w}}_i$ and $\bar{\mathbf{w}}_j$, the server needs $2(K+T+A)-1$ symbols of size a single field element from the users, for $i,j\in[N], i\ne j$. 
 Thus, the server communication load in \texttt{ByzSecAgg} is $R_{\text{server}}\le\big(1+\frac{2A+T}{K}\big)L+(T+A+K-\frac{1}{2})N(N-1)$, where the inequality is a result of the presence of dropped-out users in the setting.
	
In \texttt{ByzSecAgg}, each user participates in two rounds of secret sharing. In the first round, they send one vector of size $\frac{L}{K}$ symbols and in the second round, they send one vector of size $\frac{L}{K}+(N-1)$ symbols to each of the other users. Additionally, each user sends one vector of size $\frac{L}{K}$ symbols to the server.  However, if $K=1$,  in second round of secret sharing,  the shares from the scalar polynomial function in \eqref{share2-noise} are only sent.
 In addition, each user sends at most $\frac{N(N-1)}{2}$ pair-wise distances of size a single field element to the server. Thus, the per-user communication load in \texttt{ByzSecAgg} is upper-bounded as $R_{\text{user}}\le \min(\frac{2N}{K},N)L+\frac{3N(N-1)}{2}$. In addition, the commitment size for each user is only $3T+1$ symbols when $K=1$, and $3K+4T-2$ symbols when $K>1$, where each symbol has the size of a single group element.
 
 Therefore, the communication loads $R_{\text{server}}$ and $R_{\text{user}}$ in Theorem~1 are achieved. Additionally, according to discussions in Subsection \ref{analysis}  to meet all requirements, the inequality 
  \begin{align}
      N\ge 2A+D+\max(2K+2T-1,m+3),
  \end{align}
 must be fulfilled. In addition,  the following condition
\begin{align}
    1\le K\le \frac{N-D+1}{2}-A-T,
\end{align}
 must be satisfied for the number of partitions of the local updates, where $K\in \mathbb{Z}$.
 \end{proof}

\subsection{Computational Complexity of \texttt{ByzSecAgg}}\label{complexity}
  \textbf{Computational Complexity at the User:}
   In \texttt{ByzSecAgg}, each user $n$ performs the following three operations:\\
   1) Secret Sharing: The user $n$ generates secret shares $\mathbf{s}_{n,\tilde{n}}$ and $\tilde{\mathbf{s}}_{n,\tilde{n}}$  for $\tilde{n}\in[N]$. Generating secret shares is equivalent to the evaluation of the polynomials in \eqref{Share1} and \eqref{share2} in $N$ distinct values. If the field supports FFT, evaluation of a polynomial function of degree $m$ at $m$ points has a computational complexity of $\mathcal{O}(m\log^2m)$ \cite{kedlaya2011fast}. In \texttt{ByzSecAgg}, we need to compute polynomials of degree $K+T-1$ at $N$ points, where the coefficients
are vectors of size $\frac{L}{K}$ and $K+T-1<N$. Therefore, generating the shares has a computational cost of $\mathcal{O}(\frac{2NL}{K}\log^2N)$. In addition, user $n$ generates shares $N_{n,\tilde{n}}^{(j)}$ from scalar polynomial function in \eqref{share2-noise} for $\tilde{n}\in[N]$ and $j\in[N]\backslash\{n\}$ which has a computation cost of $\mathcal{O}(N^2\log^2N)$.\\
   2) Inner Product Computation: The user computes the inner product of the shares,  $\bar{d}_{i,j}^{(n)}$,  as defined in \eqref{distances} for $i,j\in[N], i<j$. This step has a computational complexity of $\mathcal{O}(N^2\frac{L}{K})$.\\
   3) Aggregation: The user aggregates the shares of other users belonging to the set $\mathcal{U}_{s}$.  Assuming the cardinality of $\mathcal{U}_{s}$ is $\mathcal{O}(N)$, the aggregation of secret shares for the selected users has a computational complexity of $\mathcal{O}(N\frac{L}{K})$.\\
   Therefore, the overall computational cost for each user is $\mathcal{O}(\frac{2NL}{K}\log^2N+N^2\frac{L}{K}+N^2\log^2N)$.\\
   \textbf{Computational Complexity at the Server:}
   In \texttt{ByzSecAgg}, the server performs the following three operations:\\
   1) Distance Recovery: The server should recover the pairwise distances between local updates. The computation performed by the server includes interpolation of a polynomial $\bar{d}_{i,j}(x), i,j\in[N]$ in \eqref{d_ij} of degree $2(K+T-1)<N$ using Reed-Solomon decoding. The complexity of interpolation of a polynomial
of degree $m$ is $\mathcal{O}(m\log^2m)$, when the field supports FFT \cite{kedlaya2011fast}. Thus, recovering $\mathcal{O}(N^2)$ pairwise distances has the computation cost of $\mathcal{O}(N^3\log^2N)$.\\
2) Outlier Detection: The server utilizes the Multi-Krum algorithm to eliminate outliers and select a set of users. Based on the recovered pairwise distances, the server calculates the score of each local update and selects  $m$ users with the lowest scores. This involves computing the sum of distances of each local update to its closest $N-A-2$ neighbors.  Selecting the smallest $N-A-2$ distances out of $N-1$ can be done using a sorting algorithm which has a computation cost of $\mathcal{O}(N\log N)$ per vector.  For $N$ vectors, the total complexity is $\mathcal{O}(N^2 \log N)$. Similarly, selecting $m<N-2A-2$ users with the smallest sum of distances has a cost of $\mathcal{O}(N\log N)$. In total, the complexity of this step is $\mathcal{O}(N^2\log N)$.\\
3) Recovering the Aggregation: The server interpolates polynomial function $\mathbf{F}(x)$ in \eqref{final_poly} to recover the final aggregation result of the selected users. The computational complexity of this step is $\mathcal{O}(\frac{NL}{K}\log^2 N)$.\\
Therefore, the computational cost of the server in \texttt{ByzSecAgg} is $\mathcal{O}((N^3+\frac{NL}{K})\log^2 N)$.\\

 \subsection{The Size of Finite Field in \texttt{ByzSecAgg}}
In this subsection, we analyze the minimum field size required to achieve a certain accuracy for \texttt{ByzSecAgg} to work properly. Let us assume that each user $n$ has local update $\mathbf{w}_n$, which is a \textbf{vector} of length $L$.  We use the quantization function in (1) with a quantization level $q\ge 1$ and quantization error in $[0,\frac{1}{q})$.\\
Assume that the value of the entries of the local models is bounded, i.e., $\forall n\in [N], i\in[L]$, we have~$-\tau <  [\mathbf{w}_n]_{i} < \tau$, where $\tau$ is a positive integer.  Here, $[\mathbf{w}_n]_{i}$ represents the $i$-th entry of local model $\mathbf{w}_n$. Consequently, $\forall n\in [N], i\in[L]$, $ -\tau q \le\lfloor q[\mathbf{w}_{n}]_{i}\rfloor\le \tau q-1$.
     In \texttt{ByzSecAgg}, there are two main computations: the pairwise distances and the aggregation of the local updates. These computations are performed in a finite field, which must be large enough to avoid boundary issues during the process. We choose a finite field $GF(p)$, denoted by $\mathbb{F}_p$, for some prime number $p$. The choice of $p$ must satisfy the following conditions:\\
     \textbf{Aggregation Recovery Condition:}
     To ensure the correct recovery of the aggregation, we have
     \begin{align*}
         \sum_{n\in\mathcal{U}_s} \bar{\mathbf{w}}_n &= \frac{1}{q} \Lambda^{-1}\bigg(\sum_{n\in \mathcal{U}_s} \Lambda\big(q Q(\mathbf{w}_n)\big)\bigg)\\
         &\overset{\text{(eq 1)}}{=} \frac{1}{q} \Lambda^{-1} \bigg( q \Lambda \big( \sum_{n\in \mathcal{U}_s} Q(\mathbf{w}_n)\big)\bigg)\\
         &\overset{\text{(eq 2)}}{=} \sum_{n\in\mathcal{U}_s} Q(\mathbf{w}_n)= \sum_{n\in\mathcal{U}_s} \frac{\lfloor q\mathbf{w}_n\rfloor}{q},
     \end{align*}
    where (eq 1) holds due to the linearity of the function $\Lambda(.)$ and (eq 2) holds if the following condition is met 
        $ q |\sum_{n\in \mathcal{U}_s} Q(\mathbf{w}_n)| < \frac{p-1}{2}$.
     Thus, the first condition for $p$ is
       $  p> 2N\tau q +1$.\\
     \textbf{Pairwise Distance Recovery Condition:}
     To ensure the correct recovery of the pairwise distances, we have
     \begin{align*}
         \norm{\bar{\mathbf{w}}_i-\bar{\mathbf{w}}_j} &=\frac{1}{q^2}\Lambda^{-1}\bigg( \norm{\Lambda\big(qQ(\mathbf{w}_i)\big)-\Lambda\big(qQ(\mathbf{w}_j)\big)}^2  \bigg)\\
         &=\frac{1}{q^2}\Lambda^{-1}\bigg(q^2 \norm{Q(\mathbf{w}_i)-Q(\mathbf{w}_j)}^2 \bigg)\\
         &= \norm{Q(\mathbf{w}_i)-Q(\mathbf{w}_j)}^2,
     \end{align*}
    which holds if
         $q^2 \norm{Q(\mathbf{w}_i)-Q(\mathbf{w}_j)}^2<\frac{p-1}{2}$.
     Thus, the second condition for $p$ is
        $ p> 2L (2\tau q-1)^2 +1$.\\
     Combining these conditions, we select $p$ such that
     \begin{align}\label{filed-size}
         p>2\max\{L (2\tau q-1)^2,N\tau q\}+1.
     \end{align}
  With a larger quantization level 
$q$, the scheme achieves better accuracy because the variance introduced by the quantization function decreases as $q$ increases. However, for a higher quantization level, a larger field size is required. 

       \begin{remark}
           In \texttt{ByzSecAgg}, choosing the finite field  $\mathbb{F}_p$ to satisfy \eqref{filed-size} is not a consequence of using the ramp secret sharing scheme. Since in the process of partitioning the local updates into $K$ parts, we partition a vector, not a symbol.  If $K\not| L$, we can zero-pad the local model vector. Shamir's secret sharing would require the same field size. Therefore, \texttt{BREA} would also use the same field size for a certain level of precision in the quantization step.
       \end{remark}

 \subsection{Comparison with \texttt{BREA} \cite{so2020byzantine}}
  The comparison results between \texttt{ByzSecAgg} and \texttt{BREA} \cite{so2020byzantine} are summarized in Table \ref{table}, which highlights the differences between the two schemes based on the defined parameters. In this subsection,  to demonstrate the communication load reduction performance of \texttt{ByzSecAgg} compared to \texttt{BREA}, the theoretical results presented in Table \ref{table} are evaluated for different parameter values. In addition, to have better intuition, the parameter values are selected based on  real-world and well-known architectures.

 Figure 1 compares the communication loads, measured in symbols, between \texttt{ByzSecAgg} and \texttt{BREA}. The parameter values selected for the evaluation involves $N=1000$ users, with $T=0.1N$ colluding users, $A=0.1N$ Byzantine adversaries, and $D=0.2N$ potential dropouts. The comparison focuses on three metrics: (a) per-user communication load, (b) server communication load, and (c) commitment size of each user. The number of parameters of commonly used neural networks, including GoogleNet \cite{szegedy2015going}, ResNet \cite{he2016deep}, and AlexNet \cite{krizhevsky2017imagenet}, are considered for parameter value $L$. 
 The results demonstrate that \texttt{ByzSecAgg} achieves an order-wise reduction in communication loads and commitment sizes compared to \texttt{BREA}. In this comparison, for \texttt{ByzSecAgg}, the number of partitions ($K$) is selected to minimize the aggregation of per-user communication load and server communication load. As is shown, the commitment size of the proposed scheme remains constant and is not influenced by the length of local updates. The slight variations observed in the graph are due to changes in the selected value for $K$.

 \begin{figure*}[ht!]
   \centering
  \begin{tabular}{ccc}
    \includegraphics[draft=false,width=0.3\linewidth,trim = 1.5cm 6.7cm 2cm 6.7cm, clip]{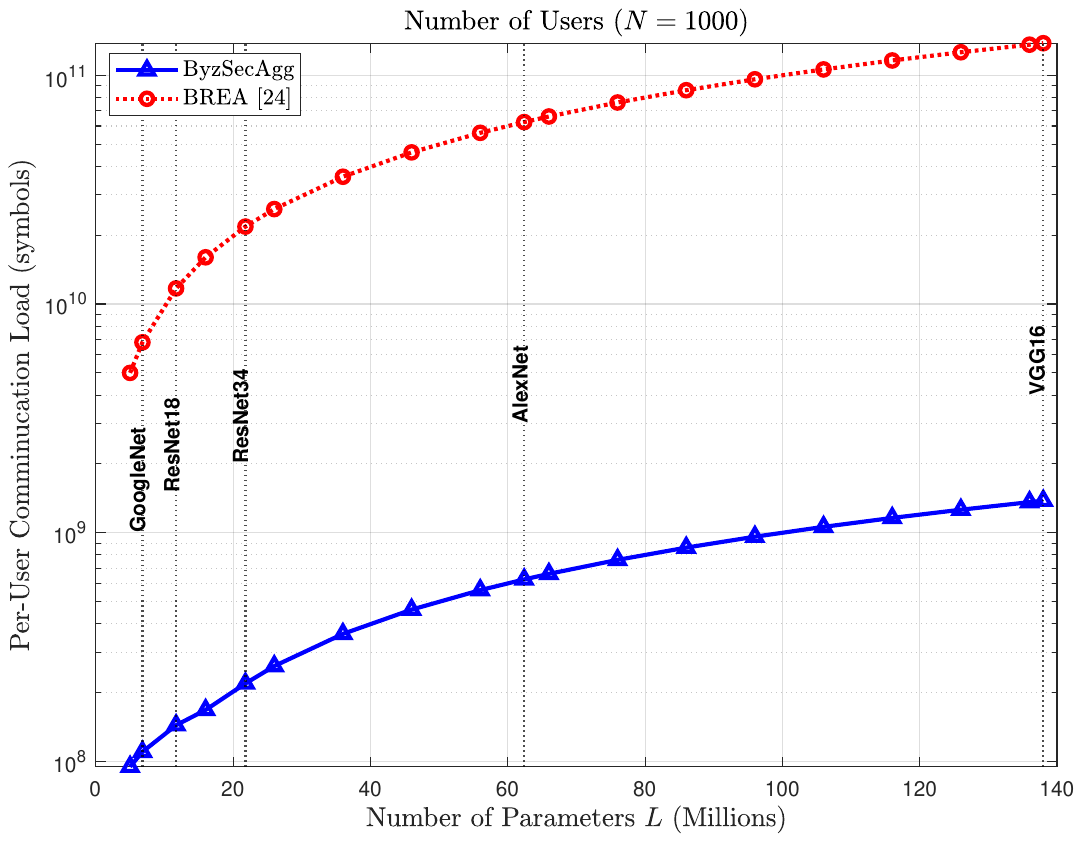} &
    \includegraphics[draft=false,width=0.3\linewidth,trim = 1.5cm 6.7cm 2cm 6.7cm, clip]{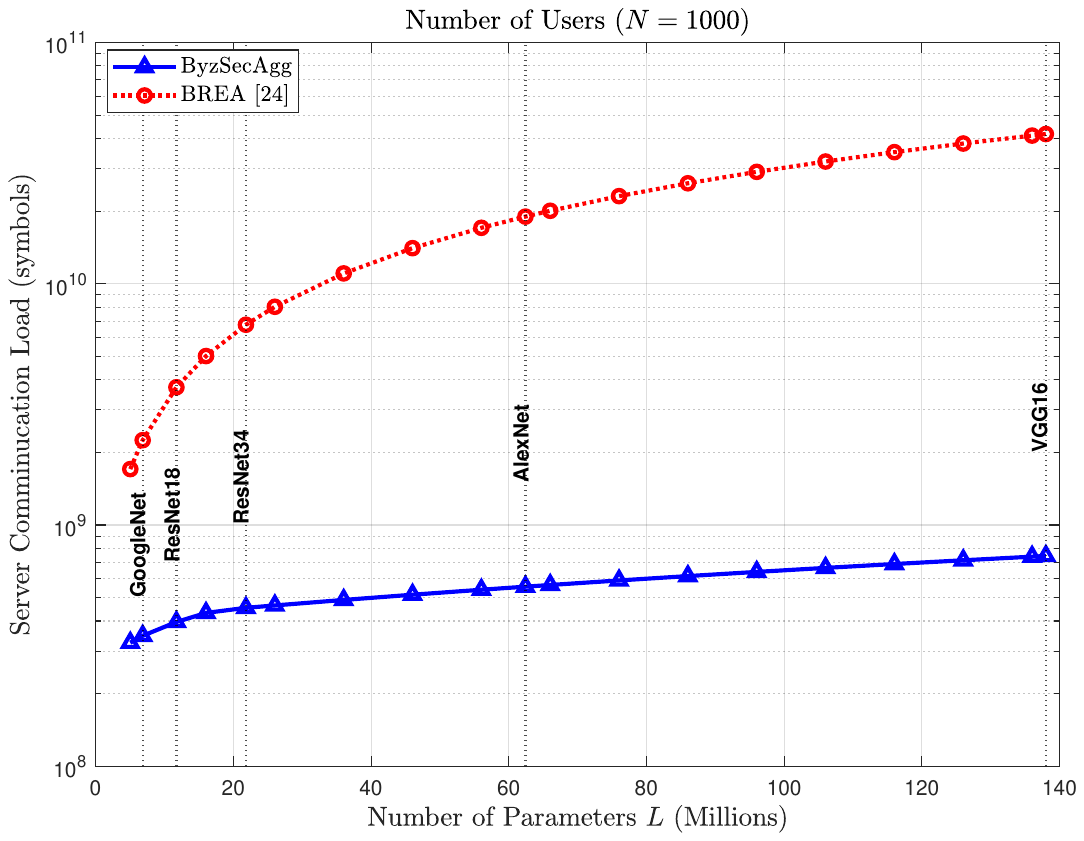} &
    \includegraphics[draft=false,width=0.3\linewidth,trim = 1.5cm 6.7cm 2cm 6.7cm, clip]{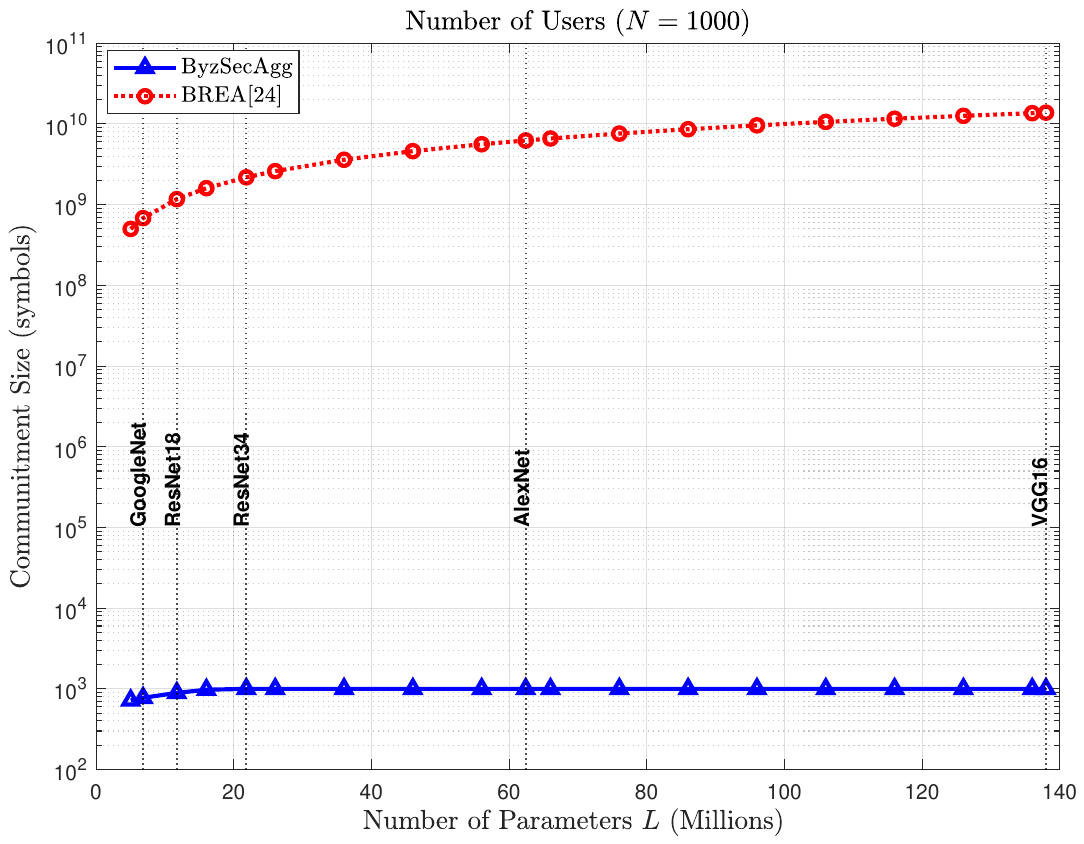} \\
    (a) & (b) & (c)
  \end{tabular}
  \caption{The comparison of communication loads, measured in symbols, between \texttt{ByzSecAgg} and \texttt{BREA} \cite{so2020byzantine} with a varying number of local update parameters ($L$). The figure illustrates the comparison using $N=1000$ users, including $T=0.1N$ colluding users, $A=0.1N$ Byzantine adversaries, and $D=0.2N$ potential dropouts. Three metrics are considered: (a) per-user communication load, (b) server communication load, and (c) commitment size of each user.}
  \label{tab:plots}
 \end{figure*}

 \begin{figure*}[ht!]
   \centering
  \begin{tabular}{ccc}
    \includegraphics[draft=false,width=0.3\linewidth,trim = 1.5cm 7cm 2cm 7cm, clip]{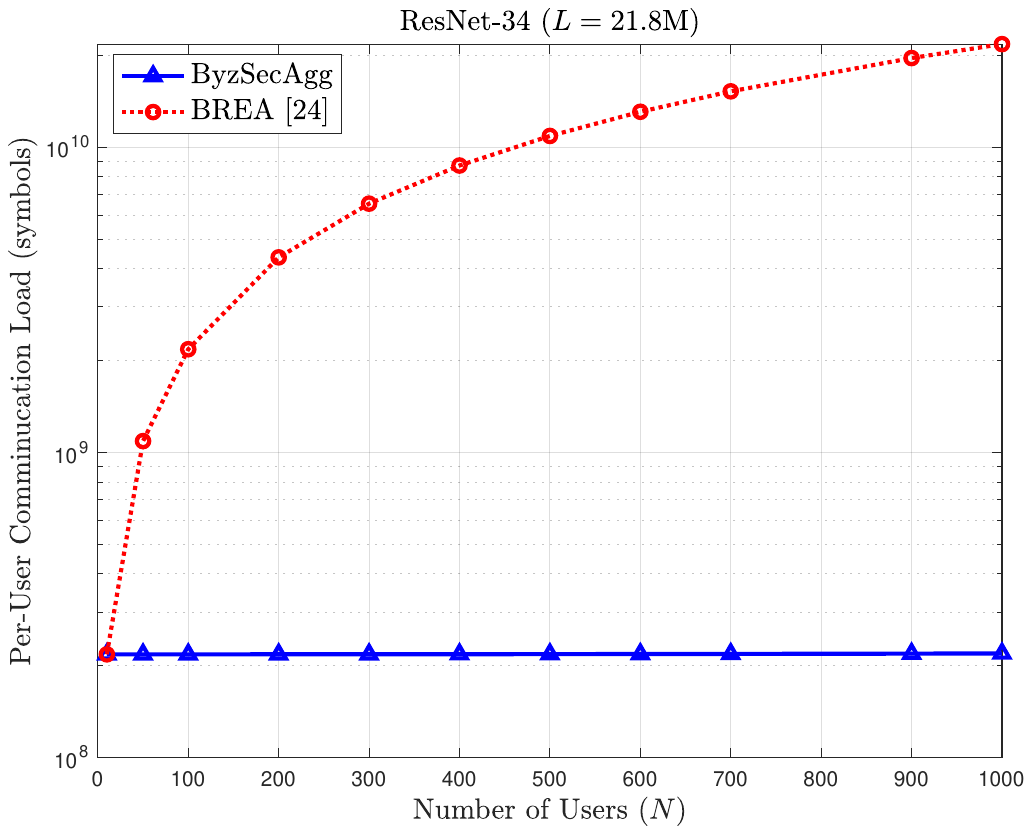} &
    \includegraphics[draft=false,width=0.3\linewidth,trim = 1.5cm 7cm 2cm 7cm, clip]{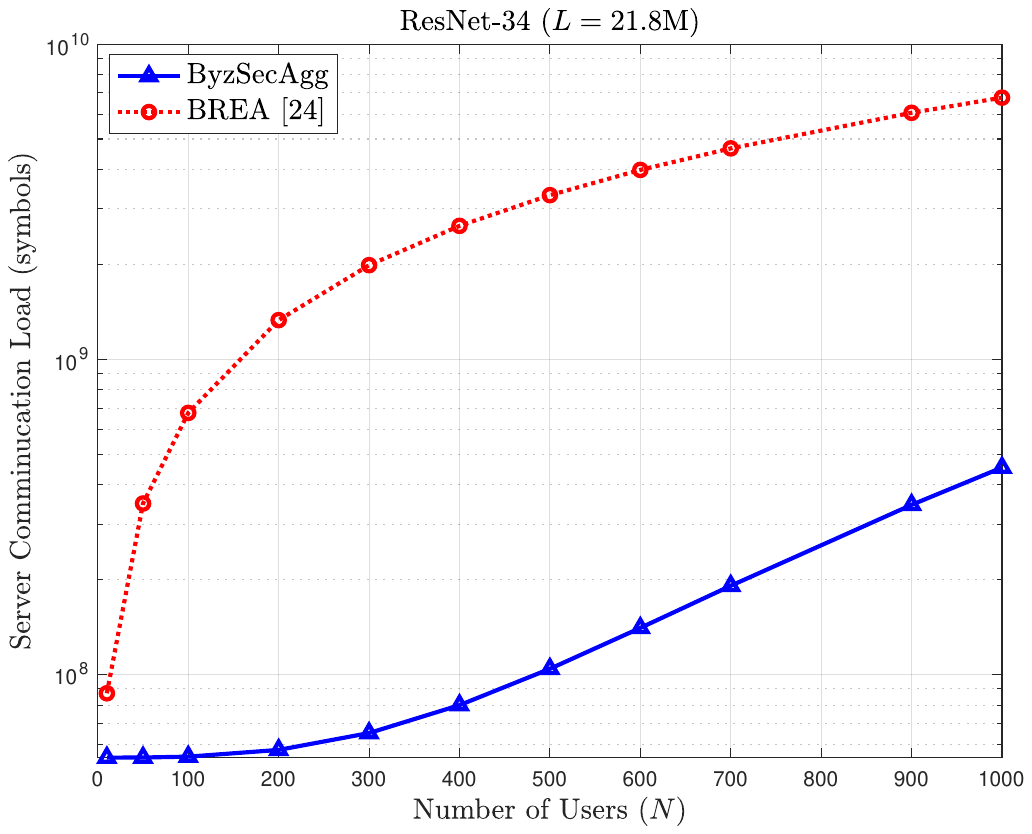} &
    \includegraphics[draft=false,width=0.3\linewidth,trim = 1.5cm 7cm 2cm 7cm, clip]{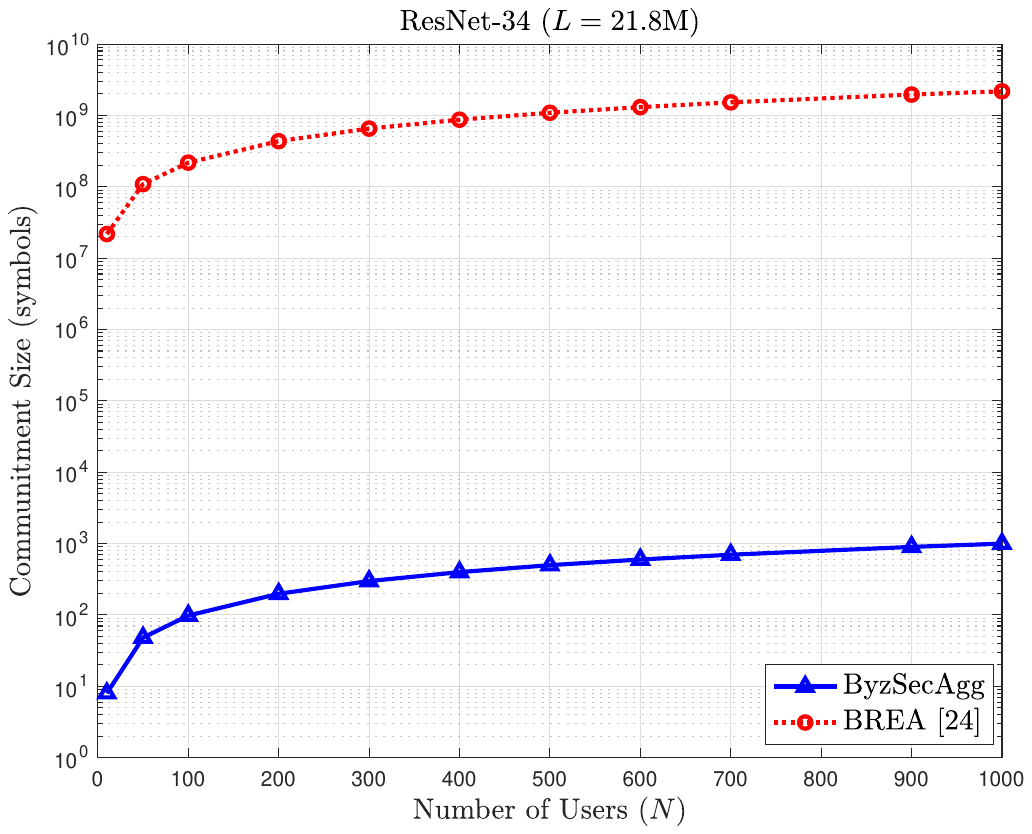} \\
    (a) & (b) & (c)
  \end{tabular}
  \caption{The comparison of \texttt{ByzSecAgg} and \texttt{BREA} \cite{so2020byzantine} in terms of communication loads while varying the number of users. The experiment considers ResNet34 with $L=21.8M$ parameters. Out of the total users ($N$), $T=0.1N$ are colluding users, $A=0.1N$ are Byzantine adversaries, and $D=0.2N$ may drop out. The comparison evaluates three metrics: (a) per-user communication load, (b) server communication load, and (c) commitment size of each user.}
  \label{tab:plots}
 \end{figure*}

Figure 2 illustrates the communication loads of the two schemes as the number of users ($N$) varies. For the evaluation, the values $L=21.8M$, $T=0.1N$ colluding users, $A=0.1N$ Byzantine adversaries, and $D=0.2N$ dropouts are considered.  Similarly, in this comparison, for \texttt{ByzSecAgg} the number of partitions ($K$) is selected to minimize the aggregation of per-user communication load and server communication load. Note that compared with \texttt{BREA}, the proposed scheme achieves the same performance in terms of convergence and resilience properties.
\begin{remark}
    Note that, in terms of privacy, \texttt{BREA} in \cite{so2020byzantine} has more information leakage compared to our proposed scheme. The reason is that, in \texttt{BREA}, each user employs Shamir's secret sharing to generate shares, computes the pairwise distances between the received shares, and sends the results back to the server. It can be shown that multiplying the shares from two Shamir's secret sharing polynomials does not preserve the privacy of the product of the secrets.
 In \texttt{ByzSecAgg}, we address this issue by using  $N_n^{(j)}(x)$ and their shares, which are employed to compute noisy inner products of shares in Step 8. Therefore, \texttt{ByzSecAgg} has a higher level of privacy.
\end{remark}
According to Subsection \ref{complexity} and \cite{so2020byzantine}, Table \ref{table-complex} provides a comparison of the computational complexity for the server and per-user operations in \texttt{ByzSecAgg} and \texttt{BREA}. The partitioning step proposed in \texttt{ByzSecAgg} reduces the computational complexity of the server and per user for large $K$.  However, in the per-user complexity, there is an additional term $N^2\log^2N$, as \texttt{ByzSecAgg} aims to achieve a higher level of privacy.
\texttt{ByzSecAgg} appears more scalable due to its reduced dependency on $L$ in both server and per-user computations when partitioning is applied $K>1$. 
\begin{table*}[!htp]
   \centering
   		\caption{%
		The comparison of \texttt{ByzSecAgg} and \texttt{BREA} \cite{so2020byzantine} in terms of computational complexity }
		\label{table-complex}
   \resizebox{1.5\columnwidth}{!}
		 {
			\begin{tabular}{||c |c c||} 
				\hline
				Approach & Server computational complexity  & Per-user computational complexity    \\ [0.5ex] 
				\hline\hline
    			\texttt{BREA}~\cite{so2020byzantine} & $\mathcal{O}((N^3+NL)\log^2N)$ & $\mathcal{O}(NL\log^2N+N^2L)$  \\
				\hline
				\texttt{ByzSecAgg}  & $\mathcal{O}((N^3+\frac{NL}{K})\log^2 N)$  & $\mathcal{O}(\frac{2NL}{K}\log^2N+N^2\frac{L}{K}+N^2\log^2N)$   \\ 
				\hline
			\end{tabular}
		  }
	\end{table*}

 \section{Conclusion}
 In this paper, we propose \texttt{ByzSecAgg}, an efficient secure aggregation scheme for federated learning that mitigates Byzantine adversarial attacks while protecting the privacy of individual local updates. \texttt{ByzSecAgg} employs techniques such as ramp secret sharing and coded computing to reduce communication loads and enable secure computation of pairwise distances, which are used for distance-based outlier detection algorithms. Additionally, we use a linear commitment scheme with a constant commitment size to ensure message integrity during the protocol and protect against adversarial behaviors of Byzantine users.
In terms of communication loads, \texttt{ByzSecAgg} outperforms the baseline scheme \texttt{BREA} and offers the advantage of a constant commitment size, regardless of local update sizes. Furthermore, it achieves the same level of performance in terms of convergence and resilience properties.

	\bibliographystyle{ieeetr}
	 \bibliography{References}
\begin{IEEEbiographynophoto}{Tayyebeh Jahani-Nezhad} received her B.Sc. degree in Electrical Engineering and M.Sc. degree in Communication Systems from Isfahan University of Technology, Iran, in 2015 and 2017, respectively. She earned her Ph.D. in Communication Systems from Sharif University of Technology, Iran, in 2022. She is currently a Postdoctoral Researcher at the Communications and Information Theory Chair (CommIT), Technische Universit\"at Berlin, Germany. Her research focuses on developing efficient and secure distributed learning frameworks, particularly in coded distributed computing and federated learning.
\end{IEEEbiographynophoto}

\begin{IEEEbiographynophoto}{Mohammad Ali Maddah-Ali} (IEEE Fellow, 2023) is an Associate Professor at the University of Minnesota Twin Cities. He received his B.Sc. degree in Electrical Engineering from Isfahan University of Technology, his M.A.Sc. degree from the University of Tehran, and his Ph.D. in Electrical and Computer Engineering from the University of Waterloo, Canada, in 2007.

From 2007 to 2008, he was with the Wireless Technology Laboratories at Nortel Networks, Ottawa, ON, Canada. He then held a Postdoctoral Fellowship at the Department of Electrical Engineering and Computer Sciences, University of California, Berkeley, from 2008 to 2010. From September 2010 to September 2020, he served as a Communication Research Scientist at Nokia Bell Labs, NJ, USA.

Dr. Maddah-Ali is the recipient of several honors, including the NSERC Postdoctoral Fellowship (2007), the Best Paper Award at the IEEE International Conference on Communications (ICC) in 2014, the IEEE Communications Society and IEEE Information Theory Society Joint Paper Award in 2015, and the IEEE Information Theory Society Paper Award in 2016. He served as an Associate Editor for the IEEE Transactions on Information Theory (2019–2022) and as Lead Editor for the IEEE Journal on Selected Areas in Information Theory. He is currently a distinguished lecturer of the IEEE Information Theory Society.
\end{IEEEbiographynophoto}

\begin{IEEEbiographynophoto}{Giuseppe Caire} (IEEE Fellow)  was born in Torino in 1965. He received a B.Sc. in Electrical Engineering  from Politecnico di Torino in 1990, an M.Sc. in Electrical Engineering from Princeton University in 1992, and a Ph.D. from Politecnico di Torino in 1994. 
He has been a post-doctoral research fellow with the European Space Agency (ESTEC, Noordwijk, The Netherlands) in 1994-1995,
Assistant Professor in Telecommunications at the Politecnico di Torino, Associate Professor at the University of Parma, Italy, 
Professor with the Department of Mobile Communications at the Eurecom Institute,  Sophia-Antipolis, France,
a Professor of Electrical Engineering with the Viterbi School of Engineering, University of Southern California, Los Angeles, and he is currently an Alexander von Humboldt Professor with the Faculty of Electrical Engineering and Computer Science at the Technical University of Berlin, Germany.

He received the Jack Neubauer Best System Paper Award from the IEEE Vehicular Technology Society in 2003,  the IEEE Communications Society and Information Theory Society Joint Paper Award in 2004 and in 2011, 
the Okawa Research Award in 2006,   
the Alexander von Humboldt Professorship in 2014, the Vodafone Innovation Prize in 2015, an ERC Advanced Grant in 2018, 
the Leonard G. Abraham Prize for best IEEE JSAC paper in 2019, the IEEE Communications Society Edwin Howard Armstrong Achievement Award in 2020, the 2021 Leibniz Prize  of the German National Science Foundation (DFG), and the 
CTTC Technical Achievement Award of the IEEE Communications Society in 2023.  Giuseppe Caire is a Fellow of IEEE since 2005.  He has served in the Board of Governors of the IEEE Information Theory Society from 2004 to 2007,
and as officer from 2008 to 2013. He was President of the IEEE Information Theory Society in 2011. 
His main research interests are in the field of communications theory, information theory, channel and source coding
with particular focus on wireless communications. 
\end{IEEEbiographynophoto}

\vfill

\end{document}